\documentclass[twocolumn, twocolappendix]{aastex63}
\usepackage[italicdiff]{physics} %italicize the d's
\usepackage{amsmath}

\usepackage{graphicx}

\newcommand{\m}[1]{\underline{\underline{#1}}}

\usepackage[encapsulated]{CJK}
\usepackage{ucs}
\usepackage[utf8x]{inputenc}
\usepackage{url}
\newcommand{\cntext}[1]{\begin{CJK}{UTF8}{bsmi}#1\end{CJK}}

\begin{document}

%% you desire. In v6.3 you can include a footnote in the title.

\title{Code Comparison in Galaxy Scale Simulations with Resolved Supernova Feedback:\\ Lagrangian vs. Eulerian Methods}

\author[0000-0002-9235-3529]{Chia-Yu Hu (\cntext{胡家瑜})}
\affiliation{Max-Planck-Institut f\"{u}r Extraterrestrische Physik, Giessenbachstrasse 1, D-85748 Garching, Germany}
%\affiliation{Center for Computational Astrophysics, Flatiron Institute, 162 5th Ave, New York, NY 10010, USA}

\author[0000-0002-9849-877X]{Matthew C. Smith}
\affiliation{Universit\"{a}t Heidelberg, Zentrum f\"{u}r Astronomie, Institut f\"{u}r theoretische Astrophysik, Albert-Ueberle-Str. 2, 69120 Heidelberg, Germany}
\affiliation{Max-Planck-Institut f\"{u}r Astronomie, K\"{o}nigstuhl 17, D-69117 Heidelberg, Germany}
%\affiliation{Harvard-Smithsonian Center for Astrophysics, 60 Garden Street, Cambridge, MA 02138, USA}
%\affiliation{Center for Computational Astrophysics, Flatiron Institute, 162 5th Ave, New York, NY 10010, USA}

\author[0000-0001-7689-0933]{Romain Teyssier}
\affiliation{Department of Astrophysical Sciences, Princeton University, Princeton, NJ 08544, USA}
%\affiliation{Centre for Theoretical Astrophysics and Cosmology, Institute for Computational Science, University of Zurich, Switzerland}

\author[0000-0003-2630-9228]{Greg L. Bryan}
\affiliation{Center for Computational Astrophysics, Flatiron Institute, 162 5th Ave, New York, NY 10010, USA}
\affiliation{Department of Astronomy, Columbia University, Pupin Physics Laboratories, New York, NY 10027, USA}

\author{Robbert Verbeke}
\affiliation{Institute for Computational Science, University of Zurich, Winterthurerstrasse 190, CH-8057 Z\"{u}rich, Switzerland}

\author{Andrew Emerick}
\affiliation{Department of Astronomy, Columbia University, Pupin Physics Laboratories, New York, NY 10027, USA}

\author[0000-0002-6748-6821]{Rachel S. Somerville}
\affiliation{Center for Computational Astrophysics, Flatiron Institute, 162 5th Ave, New York, NY 10010, USA}

\author[0000-0001-5817-5944]{Blakesley Burkhart}
\affiliation{Department of Physics and Astronomy, Rutgers University,  136 Frelinghuysen Rd, Piscataway, NJ 08854, USA}
\affiliation{Center for Computational Astrophysics, Flatiron Institute, 162 5th Ave, New York, NY 10010, USA}

\author[0000-0001-5262-6150]{Yuan Li (\cntext{黎原})}
\affiliation{Department of Physics, University of North Texas, Denton, TX 76203, USA}

\author[0000-0002-1975-4449]{John C. Forbes}
\affiliation{Center for Computational Astrophysics, Flatiron Institute, 162 5th Ave, New York, NY 10010, USA}

\author[0000-0003-2539-8206]{Tjitske Starkenburg}
\affiliation{Center for Interdisciplinary Exploration and Research in Astrophysics (CIERA) and\\Department of Physics and Astronomy, Northwestern University, 1800 Sherman Ave, Evanston IL 60201}

\correspondingauthor{Chia-Yu Hu}
\email{cyhu.astro@gmail.com}

\begin{abstract}

We present a suite of high-resolution simulations of an isolated dwarf galaxy using 
four different hydrodynamical codes:
{\sc Gizmo}, {\sc Arepo}, {\sc Gadget}, and {\sc Ramses}.
All codes adopt the same physical model
which
includes radiative cooling,
photoelectric heating,
star formation,
and supernova (SN) feedback.
Individual SN explosions are directly resolved without resorting to sub-grid models,
eliminating one of the major uncertainties in cosmological simulations.
We find 
reasonable agreement on
the time-averaged star formation rates 
as well as the joint density-temperature distributions
between all codes.
However,
the Lagrangian codes show significantly burstier star formation,
larger supernova-driven bubbles,
and stronger galactic outflows
compared to the Eulerian code.
This is caused by the behavior in the dense, collapsing gas clouds
when the Jeans length becomes unresolved:
gas in Lagrangian codes
collapses to much higher densities
than in Eulerian codes,
as the latter is stabilized by the minimal cell size.
Therefore,
more of the gas cloud is converted to stars and
SNe are much more clustered in the Lagrangian models, 
%so that the first SNe create low-density, low-radiative loss environments for subsequent SNe,
amplifying their dynamical impact.
The differences between Lagrangian and Eulerian codes
can be reduced
by adopting a higher star formation efficiency in Eulerian codes,
which significantly enhances SN clustering in the latter. 
Adopting a zero SN delay time
reduces burstiness in all codes, resulting in vanishing outflows
as SN clustering is suppressed.

\end{abstract}
\keywords{galaxy formation; hydrodynamical simulations; stellar feedback}

\section{Introduction} \label{sec:intro}

Tremendous progress 
has been made in hydrodynamical simulation of galaxy formation
in the last decade
(see \citealp{2015ARA&A..53...51S} and references herein).
Starting with initial conditions from the cosmic microwave background,
the simulated galaxies evolving across the cosmic time
are remarkably realistic in many aspects such as their
masses, sizes, metallicities, colors, morphologies, etc..
These simulations 
provide crucial information
that is often observationally inaccessible 
and therefore
have routinely been used to study the physical processes that shape the observed galaxies.

However,
galaxy formation is fundamentally a multi-scale problem
that spans several orders of magnitude both spatially and temporally.
The computational cost of simulating the 
relevant dynamical range 
is impractically high
and will remain so in the foreseeable future.
Therefore,
phenomenological sub-grid models are required to account for the 
physical processes on unresolved scales
with several tunable parameters,
rendering their predictive power somewhat ambiguous 
(see \citealp{2017ARA&A..55...59N} and references therein).

On the other hand,
small-scale simulations have provided valuable insights from a ``bottom-up'' perspective
\citep{2015MNRAS.454..238W, 2016MNRAS.456.3432G, 2016ApJ...824...41I, 2017ApJ...841..101L, 2017MNRAS.466.1903G, 2017ApJ...846..133K, 2018ApJ...853..173K, 2020ApJ...900...61K, 2021ApJ...920...44H}.
These simulations focus on a patch of the interstellar medium (ISM) on kpc scales
and therefore can achieve much higher resolution,
following
the turbulence and gravitational collapse that shapes the multi-phase ISM
down to much smaller scales and higher densities.
More importantly,
feedback from supernova (SN) explosions can be faithfully modeled when the energy-conserving 
Sedov-Taylor phase is resolved,
which requires a spatial resolution of $\sim 4$~pc in Eulerian codes \citep{2015ApJ...802...99K,2015ApJ...809...69S} 
or a mass resolution of $\sim 10~{\rm M_\odot}$ in Lagrangian codes \citep{2019MNRAS.483.3363H, 2020MNRAS.495.1035S} under typical ISM conditions.
The difficulty in resolving SN feedback in cosmological simulations is a long-standing problem 
and has motivated a plethora of sub-grid models.
Dramatically different simulation results arise 
primarily from the differences in the adopted feedback prescriptions
while the differences in the hydrodynamical techniques only play a secondary role (e.g., \citealp{2012MNRAS.423.1726S}).
Resolving SN feedback therefore eliminates a major source of uncertainty. 
However,
the periodic boundary conditions 
adopted in ISM patch simulations 
place a fundamental limitation on their applicability 
on large scales
such as the lack of spiral waves \citep{2020MNRAS.492.1594S} 
and potentially unrealistic behavior in galactic outflows \citep{2016MNRAS.459.2311M}.

Recently,
galaxy scale simulations that can resolve individual SN feedback in an isolated galaxy have become feasible \citep{2016MNRAS.458.3528H, 2016Natur.535..523F, 2017MNRAS.471.2151H, 2018MNRAS.478..302S, 2019MNRAS.483.3363H, 2019MNRAS.482.1304E, 2021MNRAS.506.3882S, 2021MNRAS.502.5417S, 2021MNRAS.501.5597G, 2022MNRAS.509.5938H}.
They have more realistic boundary conditions than the ISM patch simulations
and can accurately capture the large-scale properties.
Dwarf galaxies,
thanks to their small sizes, 
are the primary targets for this type of simulation.
More ambitious efforts have been made to pursue
SN-resolved simulations of galaxy mergers \citep{2019ApJ...879L..18L, 2020ApJ...891....2L}
or even
SN-resolved cosmological ``zoom-in'' simulations \citep{2019MNRAS.490.4447W, 2020MNRAS.491.1656A, 2022MNRAS.513.1372G, 2022MNRAS.516.5914C}.
Therefore,
it is timely
to design and perform simulations of a ``typical'' dwarf galaxy 
that can serve as a laboratory for code comparison and numerical experiments.

In this paper,
as part of the SMAUG (Simulating Multiscale Astrophysics to Understand Galaxies) project\footnote{\url{https://www.simonsfoundation.org/flatiron/center-for-computational-astrophysics/galaxy-formation/smaug}},
we conduct simulations of an isolated dwarf galaxy using four different hydrodynamical codes with the same physical model.
The SMAUG project
aims at improving the predictive power of large-scale cosmological simulations
by developing sub-grid models 
based on small-scale simulations 
(rather than calibrated to reproduce observations).
The goal of this paper is
to investigate which predictions in our small-scale simulations are robust 
and also to understand when and how differences arise.
The uniqueness of our work 
compared to previous code comparison projects
such as {\sc Agora} \citep{2016ApJ...833..202K}
is that
we do not use sub-grid models for SN feedback,
as our resolution is able to resolve it directly.
For the sake of comparison and interpretation,
we have opted for the minimal complexity that can capture the essential physics.
Our results are therefore subject to the caveat of the neglected physics,
which we will discuss in detail in Section~\ref{sec:diss}.

This paper is organized as follows.
In Sec. \ref{sec:methods},
we introduce our numerical framework and the hydrodynamical codes.
In Sec. \ref{sec:results},
we present our simulation results,
demonstrating the striking differences between Lagrangian and Eulerian codes
in the burstiness of star formation, gas morphology, and galactic outflows
as a result of SN clustering.
In Sec. \ref{sec:diss},
we discuss the implications and limitations of our results and compare them with previous work.
In Sec. \ref{sec:sum}, 
we summarize our results.

%The aim of this work is not only to find consensus among different codes but also to understand the differences when they arise.

\section{Numerical Methods} \label{sec:methods}

\subsection{Hydrodynamical Codes}

We use four different codes for gravity and hydrodynamics:
one Eulerian code ({\sc Ramses}),
one moving mesh code ({\sc Arepo}),
and two Lagrangian codes ({\sc Gizmo} and {\sc Gadget-3}),
which we briefly describe as follows.

{\sc Gadget-3} (hereafter {\sc Gadget} for short) \citep{2005MNRAS.364.1105S} is a particle-based code.
Gravity is solved using a ``tree code'' \citep{1986Natur.324..446B}
where short-range forces are calculated pairwise
while long-range forces are approximated by the center of mass of a tree node.
Since it is designed for collisionless N-body problems,
gravitational forces at small distances are reduced by softening,
and different softening lengths can be used for different components,
such as gas, stars, and dark matter.
Hydrodynamics is solved by
the smoothed particle hydrodynamics (SPH) method \citep{1977AJ.....82.1013L}.
We adopt the {\sc SPHGal} implementation in \citet{2014MNRAS.443.1173H} 
that includes 
several improvements over traditional SPH methods; these improvements
including the pressure-energy formulation \citep{2015MNRAS.450...53H},
the Wendlend kernel \citep{2012MNRAS.425.1068D},
variable artificial viscosity and conduction \citep{2008JCoPh.22710040P, 2010MNRAS.408..669C},
and the timestep limiter \citep{2012MNRAS.419..465D}.
Both gravity and hydrodynamics solvers are Lagrangian in the sense that 
particles follow the motions of the gas, stars, and dark matter.
The initial mass carried by each particle is conserved as there is no mass fluxes between particles.

{\sc Gizmo} \citep{2015MNRAS.450...53H}
is a multi-method code built on the code {\sc Gadget}, 
featuring
the meshless Godunov method \citep{2011MNRAS.414..129G} for hydrodynamics.
We adopt the meshless finite-mass (MFM) method proposed by \citet{2015MNRAS.450...53H} in this work.
In MFM,
the simulation domain is divided into overlapping cells defined by the particle distribution 
and a kernel function with a finite support radius.
The Riemann problem is solved at the interfaces between cells.
The cells move with the gas so that the mass fluxes between cells are assumed to be zero.
{\sc Gizmo} adopts the same gravity solver as {\sc Gadget}
and is a Lagrangian code when the MFM solver is used.

{\sc Arepo} \citep{2010MNRAS.401..791S,2016MNRAS.455.1134P,2020ApJS..248...32W} uses a 
second-order accurate finite
volume scheme on an unstructured, moving mesh. The simulation domain is discretized by a Voronoi tessellation and a Riemann problem is solved at the interfaces between cells to compute fluxes.
The discrete set of mesh generating points that define the tessellation are allowed to
move with a velocity close to that of the local fluid (with small corrections to maintain
cell regularity). The method is therefore pseudo-Lagrangian, as the moving mesh tends to
minimize mass fluxes between cells with the result that they roughly maintain a constant mass.
However, these mass fluxes are non-zero (in contrast
to {\sc Gadget} and {\sc Gizmo}). In the standard usage of the code, adopted here,
a refinement (de-refinement) scheme is used to split (merge) cells in order to enforce
a constant mass resolution within a factor of two. {\sc Arepo} adopts a similar gravity 
solver to those employed in {\sc Gadget} and {\sc Gizmo}.

{\sc Ramses} \citep{2002A&A...385..337T} is an Eulerian code with octree-based adaptive mesh refinement (AMR).
The N-body solver is based on the adaptive particle-mesh method. The Poisson equation is solved using the multigrid method with Dirichlet boundary conditions at level boundaries \citep{2011JCoPh.230.4756G}. Hydrodynamics is solved using the MUSCL scheme, a second order finite volume Godunov method, and the HLLC Riemann solver.
For {\sc Ramses}, we used here a quasi-Lagrangian refinement criterion based on a target mass common to all codes in this work. We use a box size of 450 kpc with a minimum level of refinement $\ell_{\rm min}=8$ and a maximum level of refinement $\ell_{\rm max}=16$ (resp. $\ell_{\rm max}=17$) for the low (resp. high) resolution run , resulting in a maximum spatial resolution of 7 pc (resp. 3.5 pc).

The mesh in Eulerian codes is static by definition.
The quasi-Lagrangian AMR leads to an adaptive spatial resolution
down to a minimal cell size.
In contrast, 
in both moving mesh and Lagrangian codes,
the resolution elements follow the motions of the gas,
leading to a smoothly (and, in principle, infinitely) adaptive spatial resolution 
with a constant or near constant mass resolution.
Despite being pseudo-Lagrangian,
moving mesh codes share many properties with Lagrangian codes.
For convenience,
we will refer to {\sc Gadget}, {\sc Gizmo} and {\sc Arepo} 
as ``Lagrangian codes'' hereafter.

%{\sc Enzo} \citep{2014ApJS..211...19B}

\subsection{Initial Conditions}
The initial conditions are generated with the {\sc MakeDiskGalaxy} code developed in \citet{2005MNRAS.361..776S},
which consist of a rotating disk galaxy embedded in a dark matter halo.
The halo has a virial radius $R_{\rm vir}$ = 45 kpc and a virial mass $M_{\rm vir}$ = $10^{10}~{\rm M}_\odot$,
and it follows a Hernquist profile which matches an NFW \citep{1997ApJ...490..493N} profile at small radii
with the concentration parameter $c$ = 15 and the spin parameter $\lambda$ = 0.035.
The baryonic mass fraction is 0.8\%,
out of which $10^7~{\rm M}_\odot$ is in the stellar disk and $7\times 10^7~{\rm M}_\odot$ in the gaseous disk.
Both the stellar and gaseous disks follow an exponential profile a with scale-length of 1 kpc,
which makes the central gas surface density $\Sigma_{\rm gas} \sim 10~{\rm M_\odot~pc^{-2}}$.
The stellar disk has a scale-height of 1 kpc,
while the vertical profile of the gaseous disk is such that it is in vertical hydrostatic equilibrium.
The initial gas temperature is set to be $10^4$ K.
The hydrogen mass fraction is $X_{\rm H} = 0.76$.
Our initial conditions resemble the Wolf-Lundmark-Melotte (WLM) galaxy,
a nearby star-forming dwarf irregular galaxy that has multi-wavelength observations \citep{2012ApJ...750...33L, 2015Natur.525..218R, 2018AJ....156..109M}.
The simulation is run for 1 Gyr for the Lagrangian codes
and 0.5 Gyr for {\sc Ramses} due to computational cost.

For {\sc Ramses} and {\sc Arepo},
it is necessary to set up a minimal density for numerical purposes,
which is set up to be $n = 10^{-7}~{\rm cm^{-3}}$ uniformly distributed in the background.
In contrast,
there is no background gas in the halo
for {\sc Gizmo} and {\sc Gadget}.

\subsection{Numerical resolution}

For comparison purposes,
we should, ideally, adopt the same resolution in both Lagrangian and Eulerian codes.
However,
Lagrangian codes have a fixed mass resolution and an adaptive spatial resolution
while Eulerian codes are the opposite.
Due to this intrinsic difference,
there is no unique way of choosing the same resolution for both methods.
In this work,
we adopt a spatial resolution of $\Delta x = 7$~pc 
in the low-resolution model
%in our low- and high-resolution models, respectively,
for the Eulerian code {\sc Ramses}.
At this resolution,
the thermal Jeans length
%$L_J = 17~{\rm pc} (T/(\mu n))^{0.5}$
can be resolved up to $n\sim 100~{\rm cm^{-3}}$ assuming $T = 30$~K.
Therefore,
for the Lagrangian codes, 
%({\sc Gizmo}, {\sc Arepo}, and {\sc Gadget}),
we choose the gas particle mass $m_{\rm g}$ such that
the smoothing length 
%$h = 1.9~\text{pc}~m_{\rm g} / n$ 
is comparable to $\Delta x$ at $n = 100~{\rm cm^{-3}}$,
%the typical densities of giant molecular clouds,
which translates to
$m_{\rm g} = $ 100 ${\rm M_\odot}$ in the low-resolution models.
The gravitational softening 
for the baryons
(i.e., gas, pre-existing stars and the stars formed in the simulation)
is set to be $h_{\rm g} = 10$~pc.
The dark matter halo is resolved with a much coarser resolution
with the particle mass $m_{\rm dm} = 10^4 ~{\rm M_\odot}$
and the gravitational softening $h_{\rm dm} = 200$~pc.
In the high-resolution models,
we simply decrease all the spatial resolutions by a factor of 2 and the mass resolutions by 8.
%$\Delta x = 3.5$~pc and $m_{\rm g} = $ 12.5 ${\rm M_\odot}$.
Table \ref{tab:res} summarizes the resolution-related parameters.
%For Lagrangian codes ({\sc Gizmo}, {\sc Gadget} and {\sc Arepo}),

\begin{deluxetable}{cccccc}
	%\tablenum{1}
	\tablecaption{
		Numerical resolution.
	}
	\label{tab:res}
	\tablewidth{0pt}
	\tablehead{
		\colhead{resolution} &
		\colhead{$m_{\rm dm} $} &
		\colhead{$m_{\rm g} $}&
		\colhead{$\ h_{\rm dm}\ $} &
		\colhead{$\ h_{\rm g}$\ } &
		\colhead{$\ \Delta x\ $} \\
		%%%%%%%%%%%%%%%%%%%%%%%
		\colhead{}   &
		\colhead{[$10^3{\rm M}_\odot$]}   &
		\colhead{[${\rm M}_\odot$]}   &
		\colhead{[pc]}   &
		\colhead{[pc]}   &
		\colhead{[pc]}   
	}
	%\decimalcolnumbers
	\startdata
	low    &   $10$	                     &	$100 $	     &	200	&	10	&	  7	\\
	high   &   $1.25$	&	$12.5$	&	100	&	5	&	3.5	\\
	\enddata
	\tablecomments{	\\
		$m_{\rm dm} $: particle mass of dark matter.  \\
		$m_{\rm g} $: particle mass of gas.\\
		$h_{\rm dm}$: gravitational softening length for dark matter.\\
		$h_{\rm g}$: gravitational softening length for gas.\\
		$\Delta x$:  cell size.
	}
\end{deluxetable}

\begin{deluxetable*}{ll r r c l}
	%\tablenum{1}
	\tablecaption{
		Overview of simulation models
	}
	\label{tab:models}
	\tablewidth{0pt}
	\tablehead{
		\colhead{model name} &
		\colhead{code} &
		\colhead{resolution} &
		\colhead{$\epsilon_{\rm SF}$} &
		\colhead{$t_{\rm SN}$ (Myr)} &
		\colhead{additional comments} 
		%%%%%%%%%%%%%%%%%%%%%%%
	}
	%\decimalcolnumbers
	\startdata
	gizmo\_lr        &        {\sc Gizmo}    &         100 M$_\odot$    &         1\%		 &		  10    \\
	arepo\_lr        &        {\sc Arepo}     &         100 M$_\odot$    &         1\%	   &		  10      \\
	gadget\_lr       &        {\sc Gadget}    &         100 M$_\odot$    &         1\%		 &		  10  \\
	ramses\_lr       &        {\sc Ramses}     &                 7 pc     &         1\%    &      10    \\
	gizmo            &        {\sc Gizmo}     &        12.5 M$_\odot$    &         1\%		 &		  10    \\
	arepo            &        {\sc Arepo}     &        12.5 M$_\odot$    &         1\%	   &		  10    \\
	ramses           &        {\sc Ramses}    &               3.5 pc     &         1\%    &      10    \\
	gizmo\_tsn0             &        {\sc Gizmo}     &        12.5 M$_\odot$    &         1\%	   &		  0      \\
	arepo\_tsn0             &        {\sc Arepo}     &        12.5 M$_\odot$    &         1\%	   &	    0   \\
	gadget\_tsn0            &        {\sc Gadget}    &        12.5 M$_\odot$    &         1\%		 &		  0   \\
	ramses\_tsn0            &        {\sc Ramses}    &               3.5 pc     &         1\%    & 		  0        \\
	ramses\_tsn50           &        {\sc Ramses}    &               3.5 pc     &         1\%    &    	50       \\
    gizmo*           &        {\sc Gizmo}     &        12.5 M$_\odot$    &         1\%		 &	    10   &  $t_{\rm SN}=0$ until $t = 0.25$ Gyr\\
	arepo*           &        {\sc Arepo}     &        12.5 M$_\odot$    &         1\%		 &	    10   &  $t_{\rm SN}=0$ until $t = 0.25$ Gyr\\
    gadget*          &        {\sc Gadget}    &        12.5 M$_\odot$    &         1\%	   &	    10   &  $t_{\rm SN}=0$ until $t = 0.25$ Gyr\\
	gizmo\_sfe100    &        {\sc Gizmo}     &        12.5 M$_\odot$    &       100\%		 &		  10 \\
	arepo\_sfe100    &        {\sc Arepo}     &        12.5 M$_\odot$    &       100\%	   &		  10    \\
	ramses\_sfe100   &        {\sc Ramses}    &               3.5 pc     &       100\%	   &		  10     \\
	\enddata
	\tablecomments{	
	    Models with an asterisk are run with $t_{\rm SN} = 0$ 
	    for the first 0.25 Gyr to prevent the initial artificial blowout 
	    of the gaseous disk.
		$\epsilon_{\rm SF}$: star formation efficiency.
		$t_{\rm SN}$: SN delay time.
	}
\end{deluxetable*}

\subsection{Radiative Cooling}
We use the public {\sc Grackle} library \citep{2017MNRAS.466.2217S}\footnote{https://grackle.readthedocs.io/} for radiative cooling,
adopting its equilibrium cooling table without solving a chemistry network.
The gas metallicity is $Z = 0.002$ (0.1 $Z_\odot$) and is constant throughout the simulation.
In addition,
we include heating from the photoelectric effect. 
Following \citet{1994ApJ...427..822B} and \citet{2003ApJ...587..278W},
the photoelectric heating rate can be expressed as
\begin{equation}\label{eq:PEheat}
	\Gamma_{\rm PE} = 1.3\times 10^{-24} n \epsilon_{\rm PE} G_0 Z^\prime_{\rm d}  ~{\rm erg~cm^{-3}~s^{-1}},
\end{equation}
where $n$ is the hydrogen number density,
$\epsilon_{\rm PE}$ is the photoelectric heating efficiency,
$G_0$ is the radiation strength in units of the Habing field \citep{1968BAN....19..421H},
and $Z^\prime_{\rm d}$ is the dust-to-gas mass ratio relative to the Milky Way value ($\sim 1\%$).
We adopt a constant $\epsilon_{\rm PE}$ of 0.05 (see Appendix~\ref{app:PEeff}) and $Z^\prime_{\rm d} = 0.1$,
assuming a linear $Z$--$Z^\prime_{\rm d}$ relationship\footnote{Observations suggest that the linear relationship between $Z$ and $Z^\prime_{\rm d}$ breaks down at low metallicity \citep{2014A&A...563A..31R}. However, its impact on the thermal balance in the ISM can be rather small if heating from SN feedback dominates over photoelectric heating \citep{2016MNRAS.458.3528H,2017MNRAS.471.2151H}.}.
We assume $G_0$ scales linearly with the SFR surface density ($\Sigma_{\rm SFR}$),
normalized to the solar-neighborhood values of $G_0 = 1.7$ and  
$\Sigma_{\rm SFR} = 2.4\times 10^{-3}~{\rm M_\odot~yr^{-1}~kpc^{-2}}$.
Since the SFR is unknown prior to the simulation,
we use low-resolution simulations to empirically determine the photoelectric heating rate 
as a function of the galactocentric radius $R$:
\begin{equation}
\Gamma_{\rm PE} = \Gamma_{\rm PE,0}  \exp( \frac{R_{\rm c} - {\rm max}(R, R_{\rm c}) }{R_{\rm c}}),
\end{equation}
where $\Gamma_{\rm PE,0} = 2.6\times 10^{-27} {\rm erg\ s^{-1}}$ 
and
$R_{\rm c}$ = 0.4 kpc.
Our low-resolution experiments suggest that the results are insensitive to 
modest variations of $\Gamma_{\rm PE}$ (a factor of 3),
which is consistent with \citet{2017MNRAS.471.2151H} and \citet{2021MNRAS.506.3882S}.

\subsection{Star Formation}

We adopt the stochastic star formation recipe commonly used in simulations of galaxy formation.
The local SFR is
$\dot{\rho}_{\rm *} = \epsilon_{\rm SF} \rho_{\rm gas} / t_{\rm ff}$
where $\rho_{\rm gas}$ is the gas density,
$\epsilon_{\rm SF}$ is the star formation efficiency
and $t_{\rm ff} \equiv  \sqrt{ 3\pi /  (32 G \rho_{\rm gas}) }$ is the free-fall time and $G$ is the gravitational constant.
We adopt $\epsilon_{\rm SF} = 1\%$ as our fiducial choice,
but we also explore $\epsilon_{\rm SF} = 100\%$ in some models.
Instead of adopted a density threshold for star formation,
we allow
star formation only to occur when 
the thermal Jeans length
$L_{\rm J} = \sqrt{ \pi c_s^2 / (G \rho_{\rm gas} ) } < L_{\rm J,0}$
where 
$c_s$ is the sound speed
and $L_{\rm J,0}$ is a predefined star formation threshold.
We adopt $L_{\rm J,0} = \Delta x$
such that
gas is eligible for star formation only when the Jeans length becomes unresolved
as we can no longer follow the gravitational collapse faithfully \citep{1997ApJ...489L.179T}. 
For Lagrangian codes,
this star formation criterion roughly corresponds
to the density where the Jeans mass becomes unresolved \citep{1997MNRAS.288.1060B}.

Star particles are stochastically created based on the local SFR.
In the Lagrangian codes, the gas resolution element is converted into a star particle that inherits the mass of its parent. This means that all star particles in {\sc Gadget} and {\sc Gizmo} have a mass of $m_{\rm g}$ exactly, while star particles in {\sc Arepo} are guaranteed to have a mass within a factor of 2 of $m_{\rm g}$. In {\sc Ramses}, a star particle 
{with a mass of $m_{\rm g}$
is spawned by removing a gas mass of $m_{\rm g}$ from the parent cell}.

%to be 7 pc and 3.5 pc in our low- and high-resolution runs, respectively,

\subsection{Supernova Feedback}

We include feedback from core-collapse SNe.
For a typical stellar population,
there is about one SN progenitor in every $100~{\rm M}_\odot$ stellar mass.
As our star particle has a mass of $m_* \leq 100~{\rm M}_\odot$,
there is at most one SN progenitor in each star particle statistically.
Therefore,
we stochastically sample massive stars 
such that each star particle has a probability of $m_* / (100~{\rm M}_\odot)$
to be selected as an SN progenitor
with an SN delay time of $t_{\rm SN} = 10$ Myr.
Each SN injects $10^{51}$ erg of thermal energy into the surrounding ISM.
For {\sc Ramses} and {\sc Arepo},
the energy is injected into the cell where the star is located.
For {\sc Gizmo} and {\sc Gadget},
multiple cells can overlap with the same star simultaneously.
For simplicity, 
we inject energy into the nearest eight particles in a kernel weighted fashion.
For each SN event,
we record not only the ambient gas density and temperature
but also the location and time 
such that we can perform a cluster analysis in Sec. \ref{sec:SNcluster}.

\subsection{Models}

The models we run are summarized in Table~\ref{tab:models}.
For the low-resolution case,
we only run our fiducial model where $\epsilon_{\rm SF} = 1\%$ and $t_{\rm SN} = 10$~Myr.
For the high-resolution case,
we run our fiducial model,
a model with $t_{\rm SN} = 0$,
and a model with $\epsilon_{\rm SF} = 100\%$.
For {\sc Ramses},
we run an additional model with $t_{\rm SN} = 50$~Myr.
Finally,
we run a model for the Lagrangian codes ({\sc Gizmo}, {\sc Arepo}, and {\sc Gadget})
where we adopt $t_{\rm SN} = 0$ up to $t = 250$~Myr and $t_{\rm SN} = 10$~Myr afterward.
%The reasons for these choices will become clear as we discuss our results.
As we will demonstrate, 
this particular setup turns out to be our favored fiducial model for Lagrangian codes
as it prevents the artificial blowout of the disk during the initial phase.
We do not have the {\sc Gadget} runs for all the models 
(i.e., there is no \textit{gadget} and \textit{gadget\_sfe100}).

\section{Results}\label{sec:results}

\begin{figure}
	\centering
	\includegraphics[width=0.9\linewidth]{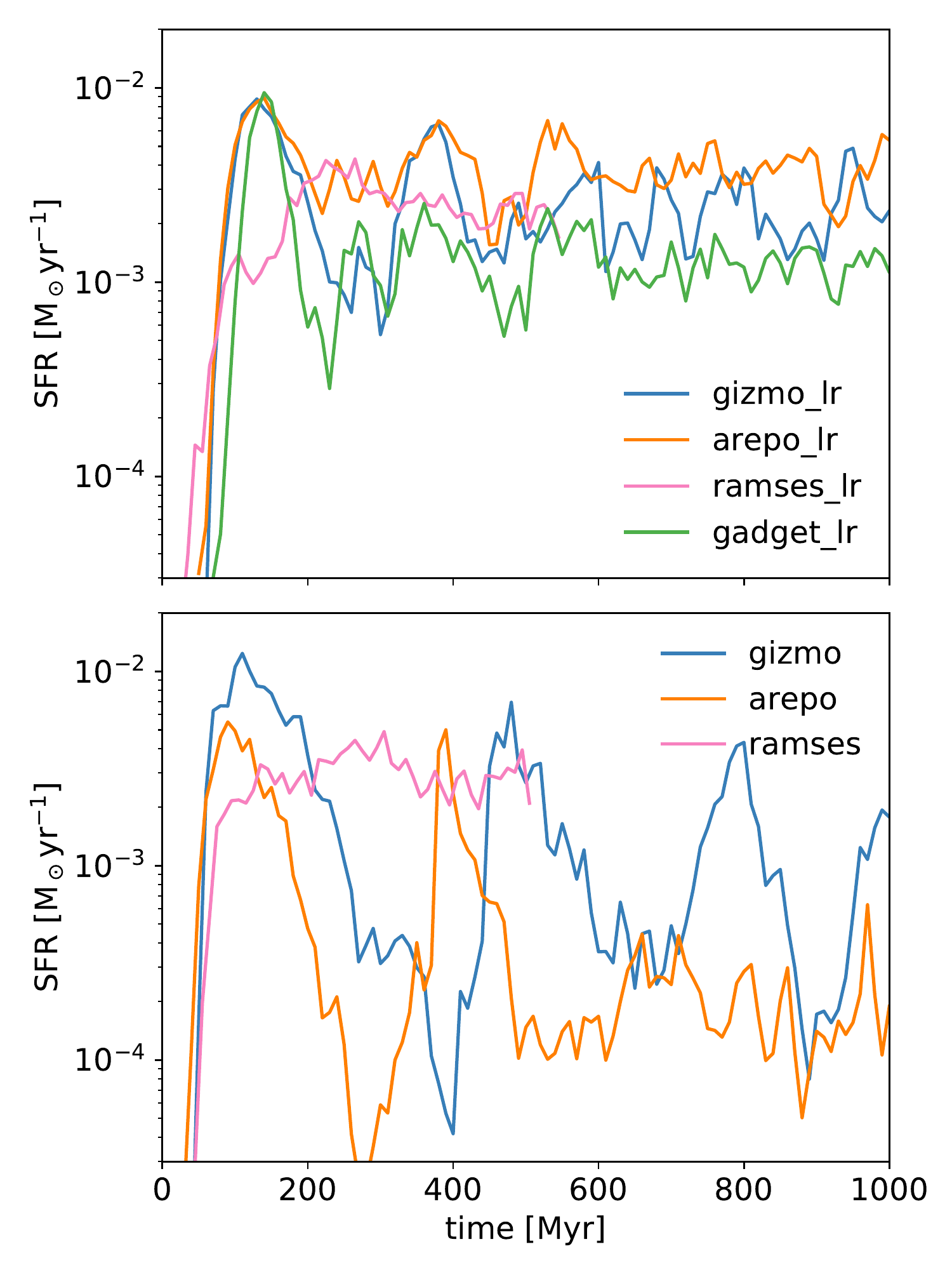}
	\caption{
		Star formation rate (SFR) as a function of time.
		Top: low-resolution models.
		%The SFR in all models are consistent within a factor of three.
		Bottom: high-resolution models (there is no corresponding {\sc Gadget} run).
		The SFR is much more bursty in the high-resolution runs in \textit{gizmo} and \textit{arepo} while it remains unchanged in \textit{ramses}.
	}
	\label{fig:sfrvstimefid}
\end{figure}

\begin{figure}
	\centering
	\includegraphics[width=0.9\linewidth]{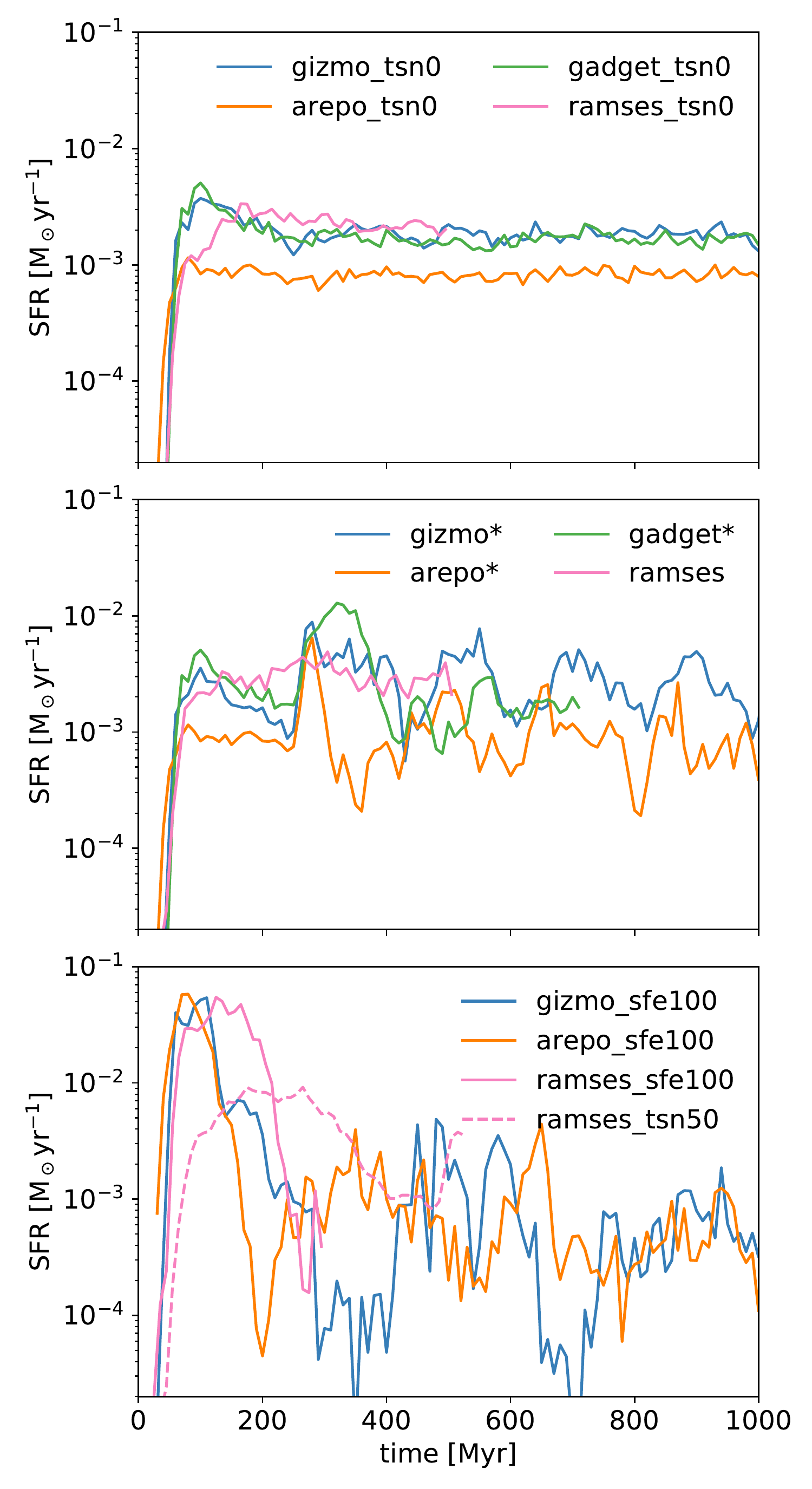}
	\caption{
		Same as Fig.~\ref{fig:sfrvstimefid} but for rest of the high-resolution models.
		\textit{Top:}
		models with instantaneous SN feedback ($t_{\rm SN}$ = 0).
		The SFR is non-bursty in all codes.
		\textit{Middle:}
		Lagrangian models
		with $t_{\rm SN}$ = 0 until $t = 250$~Myr
		(\textit{gizmo*}, \textit{arepo*} and \textit{gadget*})
		in order to mitigate the initial artificial starburst.
		%Once $t_{\rm SN}$ is set to the fiducial value of 10 Myr ,
		The Lagrangian codes are still more bursty than \textit{ramses})
		at $t \geq 250$~Myr.
		%which makes the SFR significantly less bursty (but still more bursty than \textit{ramses}).
		\textit{Bottom:}	
		models with $\epsilon_{\rm SF} = 100\%$ (solid lines) 
		and a {\sc Ramses} model with 
		$t_{\rm SN}$ = 50 Myr (\textit{ramses\_tsn50}, dashed line).			
		Increasing either $\epsilon_{\rm SF}$ or $t_{\rm SN}$ makes the SFR in {\sc Ramses} more bursty,
		similar to the Lagrangian models.
	}
	\label{fig:sfrvstimetsn0}
\end{figure}

\subsection{Star formation rate}

Fig.~\ref{fig:sfrvstimefid} shows the star formation rate (SFR) as a function of time 
for the low- and high-resolution models in the top and bottom panels, respectively.
The average SFRs in the low-resolution models are in broad agreement in all codes.
However, the \textit{ramses\_lr} model is notably less bursty 
(i.e., the SFR varies less with time) 
than the Lagrangian models.
In addition,
there is an initial starburst in the Lagrangian models at $t\sim 100$ Myr which is absent in the \textit{ramses\_lr} model.
In the high-resolution models (bottom panel),
the SFRs in the Lagrangian models become even burstier.
Indeed, 
the initial starburst leads to extremely energetic SN feedback which blows out the entire gaseous disk.
Although a quasi-steady state is eventually established after some of the blown-out gas falls back,
the overall gas surface density becomes significantly lower than that in the initial conditions,
leading to a reduced time-averaged SFR.
The blowout is, however, a somewhat artificial consequence of our initial conditions and numerical setup.
Before the first star formation occurs,
gas cools and collapses globally into an artificially thin disk.
Once the gas becomes dense enough to form stars,
the resulting coherent SNe can easily break out of the thin disk,
resulting in a catastrophic blowout.
This is a well-known phenomenon in simulations of isolated galaxies as well as in ISM patches,
and several methods have been adopted to mitigate the initial blowout,
such as turbulent driving
(e.g., \citealp{2015MNRAS.454..238W, 2017ApJ...846..133K}) or random SN driving (e.g., \citealp{2017MNRAS.471.2151H, 2021MNRAS.506.3882S}).
On the other hand,
the \textit{ramses} model
still shows a non-bursty SFR similar to its low-resolution counterpart 
and there is no initial blowout of the disk,
in sharp contrast to the Lagrangian models.
%The resolution ($\Delta x = 3.5$~pc)
%is comparable to the ISM patch simulations 
%in \citet{2015MNRAS.454..238W, 2016MNRAS.456.3432G, 2017ApJ...846..133K}

\begin{figure*}[htp]
	\centering
	\includegraphics[width=0.99\linewidth]{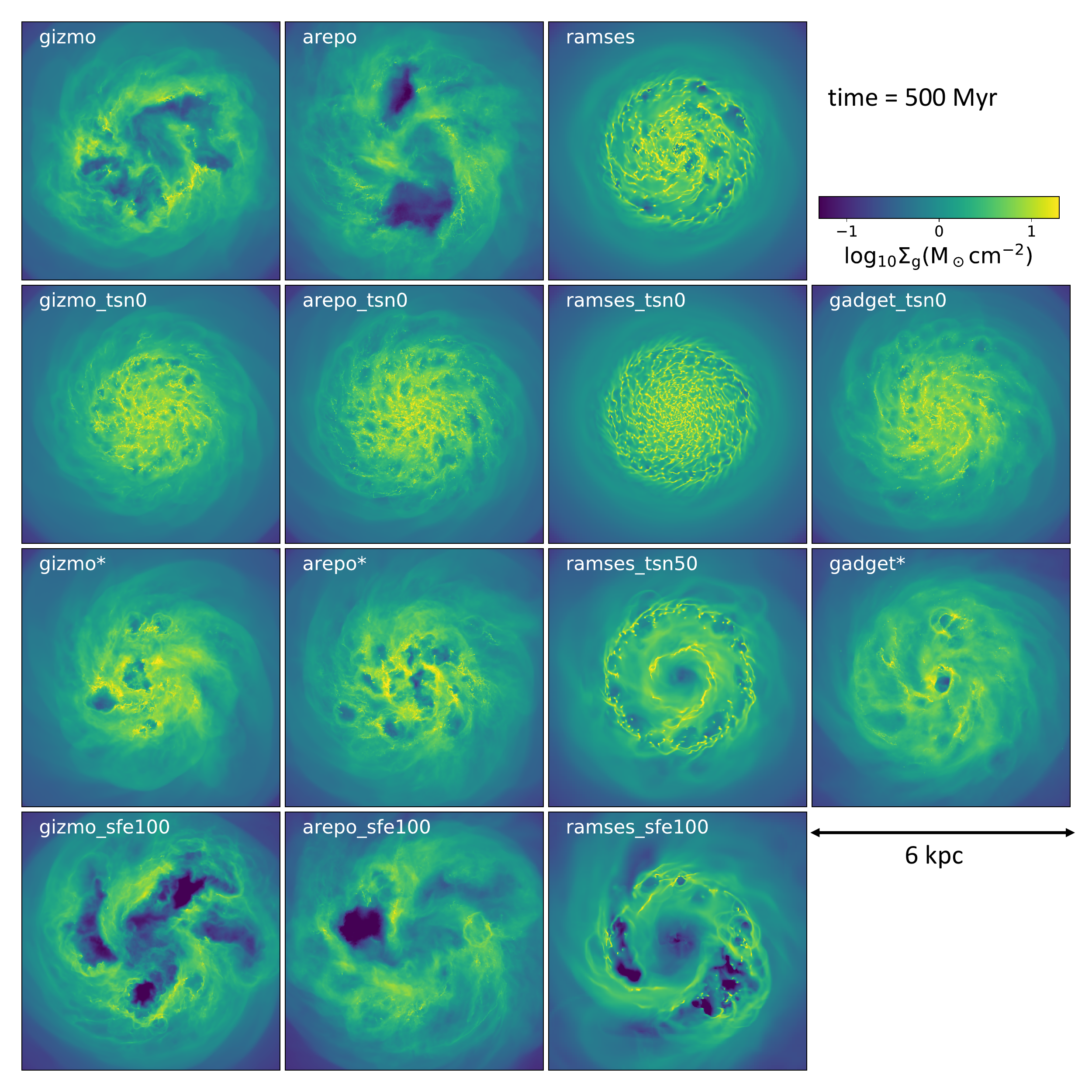}
	\caption{
		Gas surface density in the face-on view in different models at $t = 500$~Myr.	
		%Instantaneous SN feedback leads to a 
		The SN-driven bubbles are clearly visible in most cases
		except in models with $t_{\rm SN}$ = 0 and in \textit{ramses}.
	}
	\label{fig:maps_all_scale}
\end{figure*}

An important clue in understanding this difference comes from the observation that
the SN delay time has a significant effect on the burstiness of star formation
in the Lagrangian codes.
%Fig.~\ref{fig:sfrvstimetsn0} shows the SFR as a function of time
%for the other high-resolution models.
The top panel of Fig.~\ref{fig:sfrvstimetsn0} shows 
the SFR time evolution
for models with instantaneous SN feedback ($t_{\rm SN}$ = 0).
With this change, the SFRs in \textit{gizmo\_tsn0}, \textit{ramses\_tsn0}, and \textit{gadget\_tsn0} all show excellent agreement with each other.
The SFR in \textit{arepo\_tsn0} is lower by a factor of 2,
{and it is unclear what causes the slight difference.}
More importantly,
the SFRs in all codes are strikingly constant with little temporal fluctuation.
In addition,
the initial burst is absent and a quasi-steady state is rapidly established
without the blowout of the disk.

\begin{figure*}[htp]
	\centering
	\includegraphics[width=0.99\linewidth]{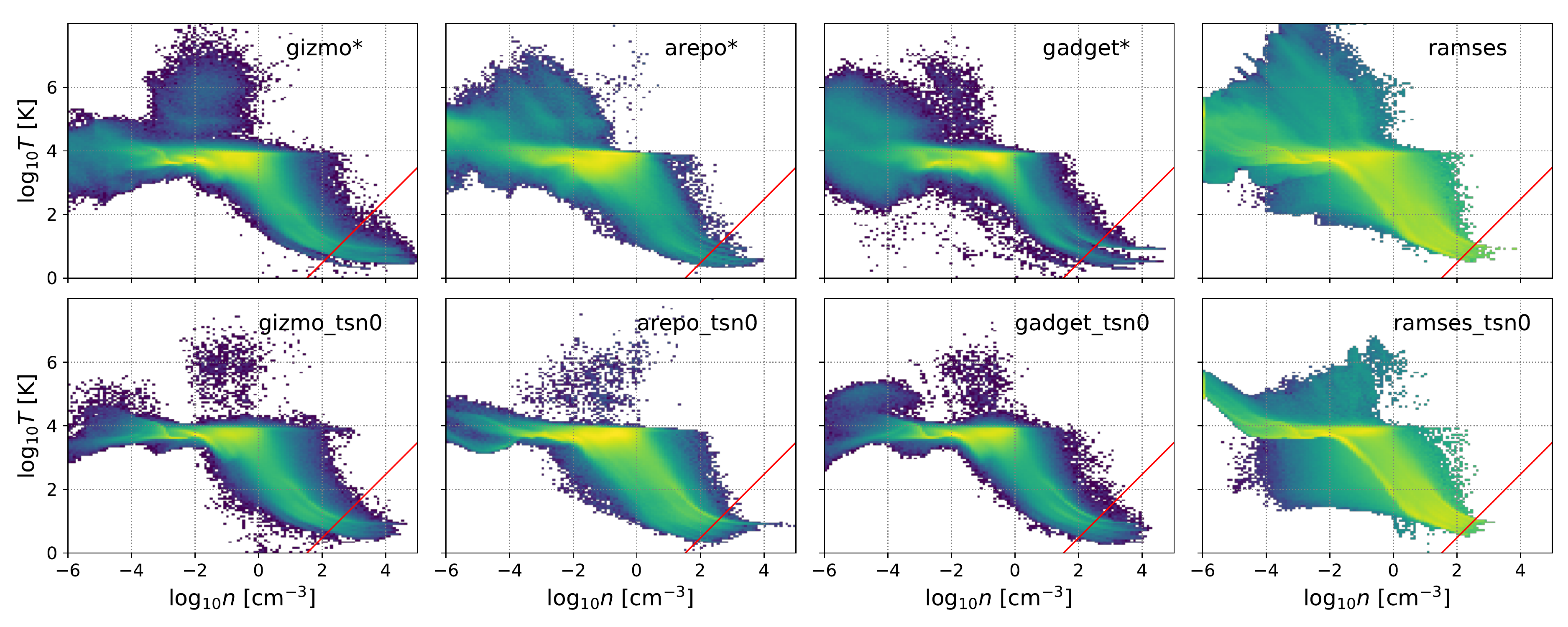}
	\caption{
		Two-dimensional {mass-weighted} normalized distribution of hydrogen number density ($n$) vs. temperature ($T$) at $t = $ 500 Myr.
		The red solid line indicates the star formation threshold where the Jeans length $L_{\rm J} = 3.5$~pc.
		All models show broadly similar distributions,
		though models with $t_{\rm SN} = 0$ produce less hot and diffuse gas compared to their fiducial counterparts.		
	}
	\label{fig:pdcodecompare}
\end{figure*}

An interesting implication is that
instantaneous SN feedback can be a simple alternative method to mitigate the artificial initial starburst.
Motivated by this, 
we rerun our Lagrangian codes
with $t_{\rm SN}$ = 0 for $t < 250$~Myr
and $t_{\rm SN}$ = 10 Myr for $t \geq 250$ Myr.
We do not do so with {\sc Ramses} as it shows no initial starburst.
The results are
shown in the middle panel of Fig.~\ref{fig:sfrvstimetsn0}.
The Lagrangian models show reasonable agreement with each other.
Although there is still a burst of SFR at $t \gtrsim 250$ Myr 
right after we switch $t_{\rm SN}$ from 0 to 10 Myr, 
it is much weaker and does not blow out the disk.
Consequently,
the time-averaged SFR is not reduced and is
comparable to \textit{ramses}.
We therefore refer to these models as our ``fiducial'' Lagrangian models.
However,
even if the artifacts of the initial conditions have been greatly reduced,
the SFRs for the Lagrangian codes are still burstier than \textit{ramses}.
This suggests that the difference in burstiness between the two methods
is an intrinsic property rather than an artifact of the initial conditions.

The observed SFR in the WLM galaxy
from \citet{2010AJ....139..447H}
is 
and $1.7\times 10^{-3}~{\rm M_\odot~yr^{-1}}$ from H$\alpha$
%(representing star formation in the past 5~Myr or so)
and
$6.3\times 10^{-3}~{\rm M_\odot~yr^{-1}}$ from far-ultraviolet (FUV). 
%(FUV, representing star formation averaging over a longer time).
The SFRs in our fiducial models are broadly in good agreement with observations,
which is reassuring.

%Having shown that the Lagrangian codes become non-bursty with $t_{\rm SN} = 0$,
%we now demonstrate how we can make the Eulerian codes bursty.
The bottom panel of Fig.~\ref{fig:sfrvstimetsn0} shows models with 
%an extremely high star formation efficiency, 
$\epsilon_{\rm SF}$ = 100\%
as well as a {\sc Ramses} model with $t_{\rm SN}$ = 50 Myr.
Both \textit{gizmo\_sfe100} and \textit{arepo\_sfe100} show very bursty SFRs
similar to the corresponding models \textit{gizmo} and \textit{arepo} where $\epsilon_{\rm SF}$ = 1\%. 
%shown in the lower panel of Fig.~\ref{fig:sfrvstimefid}).
%The burstiness in Lagrangian codes is insensitive to $\epsilon_{\rm SF}$.
On the other hand,
both \textit{ramses\_sfe100} and \textit{ramses\_tsn50} are bursty,
in contrast to \textit{ramses}.
In other words,
the burstiness in Eulerian codes
can be enhanced by 
increasing either $\epsilon_{\rm SF}$ or $t_{\rm SN}$.
We will explore the reason for this in later sections.

%Increasing $\epsilon_{\rm SF}$ 

The time-averaged SFR of different models are summarized in Table~\ref{tab:sum_table}.

\subsection{Gas morphology and thermal state}

Fig.~\ref{fig:maps_all_scale} shows the 
face-on maps of the gas surface density 
in different models at $t = $ 500 Myr.	
The most striking feature 
is that
large SN-driven bubbles 
are clearly seen in some models 
but are completely absent in others.
In fact,
the presence of SN bubbles
is closely correlated with the burstiness of the SFR 
shown in Figs.~\ref{fig:sfrvstimefid} and \ref{fig:sfrvstimetsn0}:
models with burstier star formation show larger SN bubbles.
In particular,
those that experience an initial blowout 
have the largest SN bubbles
and their gas surface densities are notably reduced even at $t = $ 500 Myr,
as some of the blown-out gas never falls back to the disk.	
On the other hand,
SN bubbles are completely absent in models where $t_{\rm SN}$ = 0.
Similarly, the standard
\textit{ramses} runs shows very few SN bubbles,
consistent with its non-bursty star formation history.
SN bubbles in {\sc Ramses} models can only be generated
by increasing either $\epsilon_{\rm SF}$ (\textit{ramses\_sfe100}) 
or $t_{\rm SN}$ (\textit{ramses\_tsn50}).

Fig.~\ref{fig:pdcodecompare} shows the 
two-dimensional {mass-weighted} normalized distribution of the hydrogen number density 
vs. temperature (the so-called ``phase diagram'') at $t = $~500 Myr.
The red solid line indicates the star formation threshold where $L_{\rm J} = 3.5$~pc.
Broadly speaking,
all models show similar distributions in the phase diagram
despite their differences in star formation burstiness and gas morphology.
This is because
the thermal balance in the ISM is mainly 
controlled by radiative cooling.
The majority of gas is in the diffuse warm gas where $n\sim 0.1\ {\rm cm^{-3}}$ and $T\sim 10^4$ K,
which is consistent with \citet{2016MNRAS.458.3528H, 2017MNRAS.471.2151H}.
However,
models with $t_{\rm SN} = 0$ produce less hot ($T > 10^5$ K) and diffuse gas
compared to their fiducial counterparts.
This is expected as 
the hot and diffuse gas typically exists in 
the SN bubbles that are largely absent in these models.
{The {\sc Ramses} runs have more gas in the density range of 
$n\sim 10 - 100\ {\rm cm^{-3}}$ as indicated by the slightly brighter color
but have very little gas above the star formation threshold.
}

%\begin{figure}
%	\centering
%	\includegraphics[width=0.99\linewidth]{gasPDF_compare}
%	\caption{}
%	\label{fig:gaspdfcompare}
%\end{figure}

\subsection{Environments for supernovae and star formation}\label{sec:nSFnSN}

%The striking difference in burstiness between different models
%can be explained by the environment where SNe occur.

\begin{figure*}[htp]
	\centering
	\includegraphics[width=0.99\linewidth]{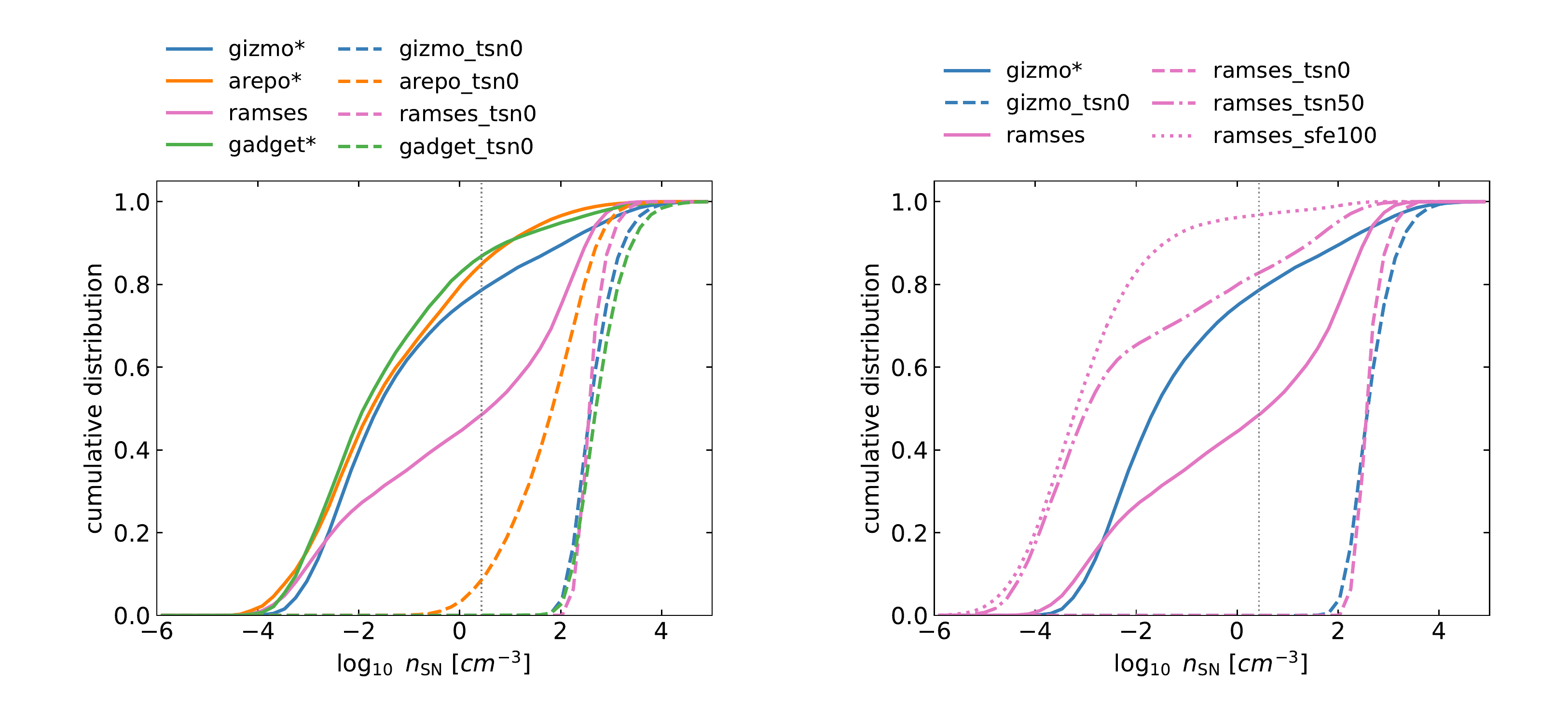}
	\caption{
		Cumulative distribution of the ambient density where SNe occur ($n_{\rm SN}$).
		{The vertical dotted line indicates the density above which the SN cooling radius becomes unresolved.}
		\textit{Left:}
		fiducial and instantaneous SN models in Lagrangian codes.
		Most SNe occur in diffuse gas where $n_{\rm SN} \sim 10^{-2}~{\rm cm^{-3}}$ in the fiducial models
		while most SNe  occur in dense gas where $n_{\rm SN} \sim 10^{2}~{\rm cm^{-3}}$ in models where $t_{\rm SN}$ = 0.
		The agreement between different Lagrangian codes is remarkable.
		\textit{Right:}
		comparison between Lagrangian ({\sc Gizmo}) and Eulerian ({\sc Ramses}) models.
		The fiducial {\sc Ramses} model (\textit{ramses}) shows a significantly lower fraction of SNe occuring in diffuse gas than \textit{gizmo*}.
		Increasing either $\epsilon_{\rm SF}$ or $t_{\rm SN}$ in {\sc Ramses} greatly increases the probability of SNe occuring in diffuse gas.
	}
	\label{fig:sncdfcompare}
\end{figure*}

The environment where SNe occur provides crucial information on the efficiency of SN feedback.
The left panel of Fig.~\ref{fig:sncdfcompare}
shows the cumulative distribution of the ambient gas density where SNe occur ($n_{\rm SN}$)
for the Lagrangian models.
The agreement between different Lagrangian codes is remarkable.
Most SNe occur in diffuse gas where $n_{\rm SN} \sim 10^{-2}~{\rm cm^{-3}}$ in the fiducial models 
(\textit{gizmo*}, \textit{arepo*}, and \textit{gadget*})
and therefore are well resolved.
In contrast,
most SNe occur in dense gas where $n_{\rm SN} \sim 10^{2}~{\rm cm^{-3}}$ in the instantaneous SN models 
(\textit{gizmo\_tsn0}, \textit{arepo\_tsn0}, and \textit{gadget\_tsn0}).
This is expected, as SN feedback kicks in right after star formation
which by construction only occurs in cold and dense gas\footnote{{We note that \textit{arepo\_tsn0} is implemented in a way that the SN delay time is very small but not exactly zero.
This explains why its cumulative distribution of $n_{\rm SN}$ is not exactly the same as that of $n_{\rm SF}$ but is instead slightly shifted to lower values,
which might explain the factor-of-two difference in the SFR in Fig.~\ref{fig:sfrvstimetsn0}.
}},
leading to more efficient energy loss from radiative cooling 
which explains the non-burstiness and the lack of SN bubbles.
Although these SNe are formally unresolved,
the time-averaged SFRs in these models are still comparable to those in the fiducial models,
suggesting that
these ``unresolved'' SNe are still able to inject enough momentum to regulate star formation.

In contrast,
\textit{ramses}
shows a very different $n_{\rm SN}$ distribution compared to the Lagrangian codes
with a significant fraction of SNe occurring at $n > 10~{\rm cm^{-3}}$,
as shown in the right panel of Fig.~\ref{fig:sncdfcompare}.
This is part of the explanation for why 
\textit{ramses} behaves more like the instantaneous SN models.
{However,
there are still $\sim 25\%$ of SNe with $n_{\rm SN} < 0.01~{\rm cm^{-3}}$ in \textit{ramses},
which is not significantly lower compared to \textit{gizmo*}.
This implies that SNe occurring at low densities is not a sufficient condition for efficient feedback,
which we will discuss in more detail in Section~\ref{sec:SNcluster}.
}
On the other hand,
both \textit{ramses\_sfe100} and \textit{ramses\_tsn50}
have a large fraction of low-$n_{\rm SN}$ SNe
and therefore show busty SFRs and large SN bubbles.

%there is a correlation between burstiness and nSN less than mean ISM density ($n<0.1$).

\begin{figure*}
	\centering
	\includegraphics[width=0.99\linewidth]{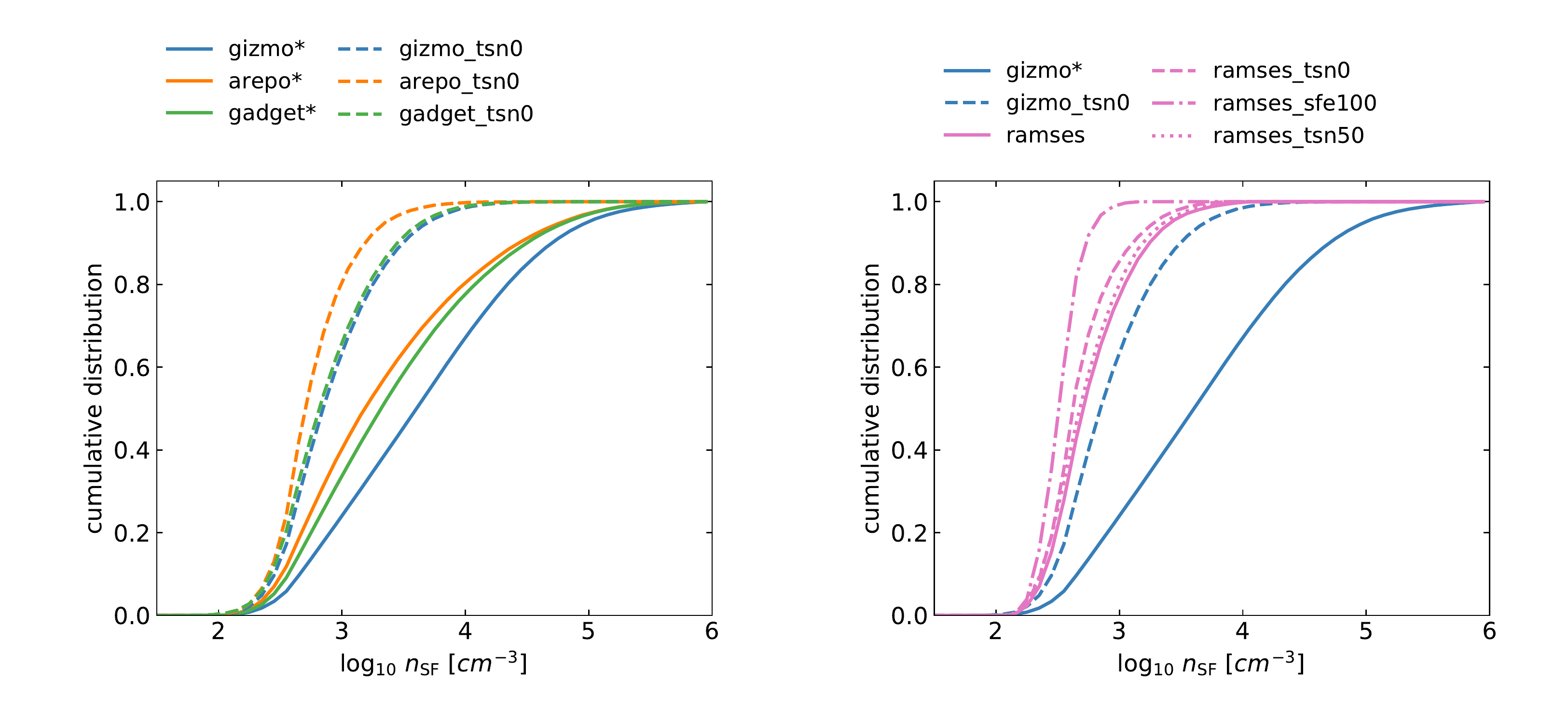}
	\caption{
		Same as Fig.~\ref{fig:sncdfcompare} but for the ambient density where star formation occurs ($n_{\rm SF}$).
		In Lagrangian codes,
		gas collapses way beyond the star formation threshold density in the fiducial models.
		In contrast,
		star formation mostly occurs around the threshold density
		in all {\sc Ramses} models.
	}
	\label{fig:sfcdf}
\end{figure*}

Having established that
the burstiness and gas morphology are both related to $n_{\rm SN}$,
we now need to understand
what leads to the difference in $n_{\rm SN}$ in different models.

The left panel of Fig.~\ref{fig:sfcdf}
shows the cumulative distribution of the gas density where star formation occur ($n_{\rm SF}$)
for the Lagrangian codes.
The agreement between different Lagrangian codes is again remarkable.
With instantaneous SN feedback (dashed lines),
most star formation occurs in a narrow range of densities 
where $n\sim 500~{\rm cm^{-3}}$.
Once the gas reaches high enough densities to form stars,
SN feedback that occurs instantaneously is able to stop the gas from further collapsing.
Therefore,
stars always form around the star formation threshold density.
In contrast,
in the fiducial models
with $t_{\rm SN} = 10$~Myr (solid lines),
star formation occurs over a large range of densities, 
spanning more than two orders of magnitude
from $500~{\rm cm^{-3}}$ to $10^5~{\rm cm^{-3}}$.
In this case,
gas keeps collapsing toward higher densities
even after it exceeds the threshold density,
as there is no countering mechanism against gravity
before the first SNe occur in 10~Myrs.
In contrast,
as shown in the right panel of Fig.~\ref{fig:sfcdf},
\textit{ramses} behaves very differently from its Lagrangian counterparts,
showing a narrow range of $n_{\rm SF}$
very similar to the instantaneous SN models.

We therefore conclude that
Lagrangian and Eulerian codes behave very differently 
at densities above the star formation threshold
where the Jeans length becomes unresolved:
%Different behavior when gas passes the SF threshold!
while gas in Lagrangian codes continues to collapse to much higher (unresolved) densities, 
%effectively boosting  $\epsilon_{\rm SF}$,
gas in Eulerian codes lingers around the star formation threshold.
This is because
the spatial resolution in Lagrangian codes is adaptive
and therefore it is much easier for the unresolved gas to collapse.
In comparison,
the collapse of unresolved gas
in Eulerian codes 
is limited by the fixed spatial resolution
and is thus relatively more difficult.
Note that in both cases, the collapse is unresolved (and additional physical processes such as pre-SN feedback may become important) and so neither behavior is necessarily correct.

%Three scenarios:
%(1) SF is locally enhanced ($\rho^{1.5}$ Schmidt law).
%(2) SNe escape their parent clouds.
%(3) SNe are clustered.

%(1) is ruled out as high nSF does not correspond to high nSN! 
%From the FOF plot,
%(3) is favored over (2).

%\begin{figure*}
%	\centering
%	\includegraphics[width=0.99\linewidth]{densSFvsSN}
%	\caption{
	%	Two-dimensional distribution of $n_{\rm SF}$ vs. $n_{\rm SN}$.
	%	The dashed line indicates $n_{\rm SF} = n_{\rm SN}$.
	%}
	%\label{fig:denssfvssn}
%\end{figure*}

\subsection{Supernova clustering}\label{sec:SNcluster}

\begin{figure*}
	\centering
	\includegraphics[width=0.99\linewidth]{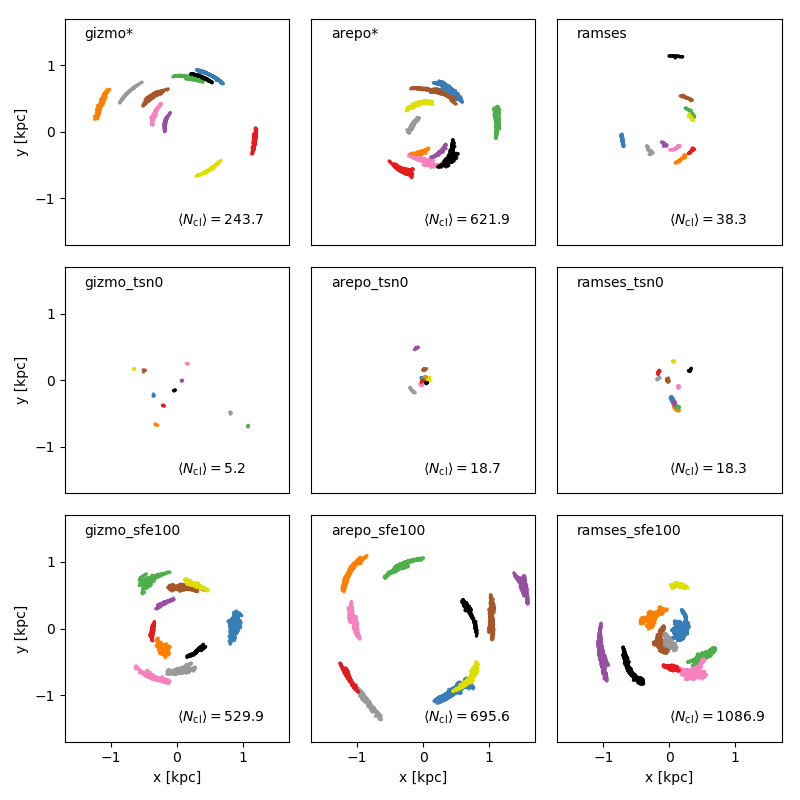}
	\caption{
		Top ten SN clusters (ranked by the clustering number $N_{\rm cl}$) in each model in the face-on view. 
		Each circle represents an SN event while each color represents an SN cluster.
		The average clustering number of the top ten clusters is shown as $\left< N_{\rm cl} \right>$.
		SNe are significantly less clustered in models with $t_{\rm SN} = 0$ as well as in the fiducial {\sc Ramses} model (\textit{ramses}).		
	}
	\label{fig:fofscatterall}
\end{figure*}

\begin{figure}
	\centering
	\includegraphics[width=0.99\linewidth]{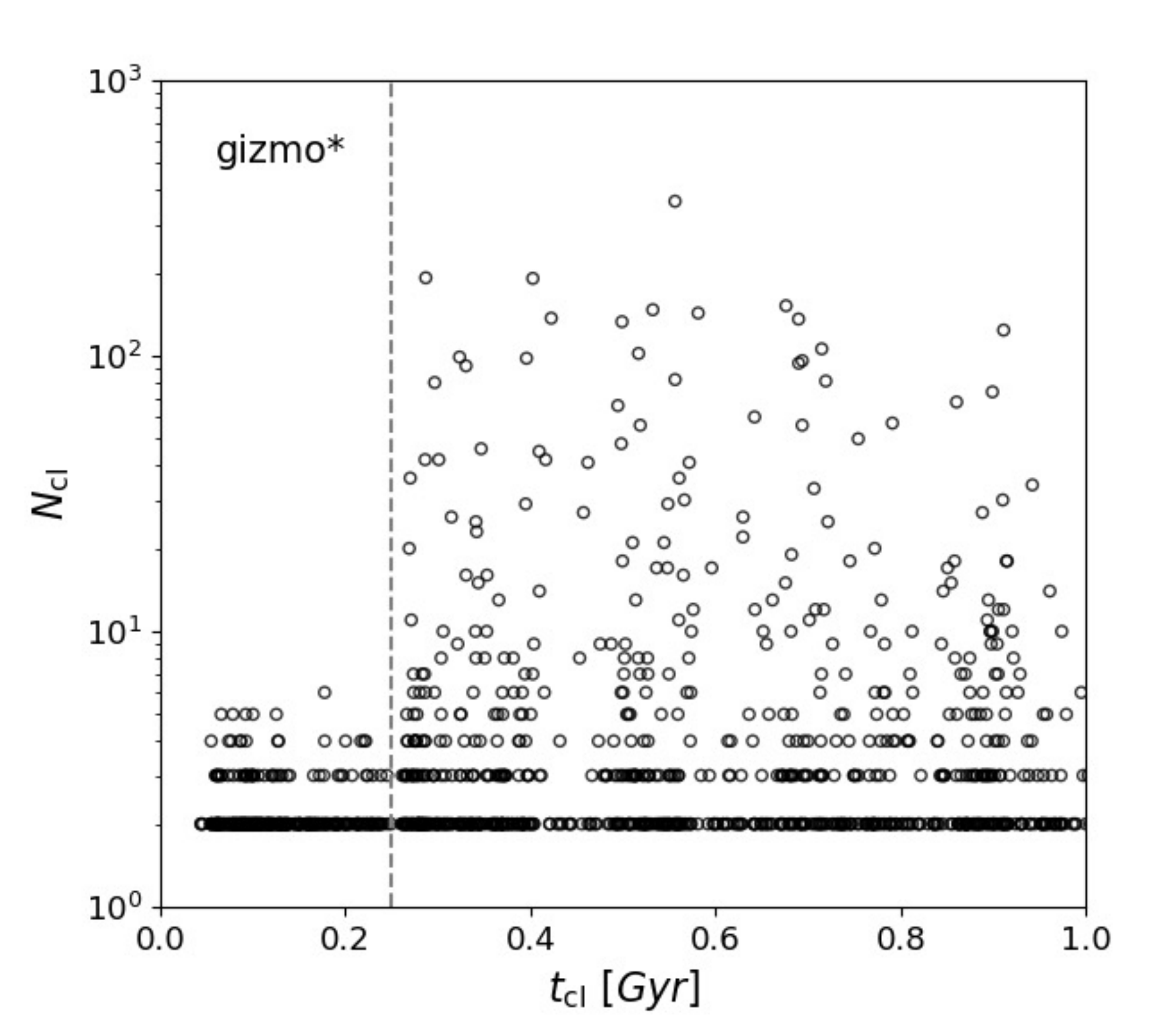}
	\caption{
		SN clustering number ($N_{\rm cl}$) as a function of the median time of the SNe in a cluster ($t_{\rm cl}$)
		in the model \textit{gizmo*}.
		Each circle represents an SN cluster.
		The vertical dashed line indicates the time where $t_{\rm SN}$ is switched from 0 to 10 Myr, 
		which leads to a substantial increase in $N_{\rm cl}$.
	}
	\label{fig:foftime}
\end{figure}

\begin{figure*}
	\centering
	\includegraphics[trim=1cm 1cm 1cm 1cm,clip, width=0.99\linewidth]{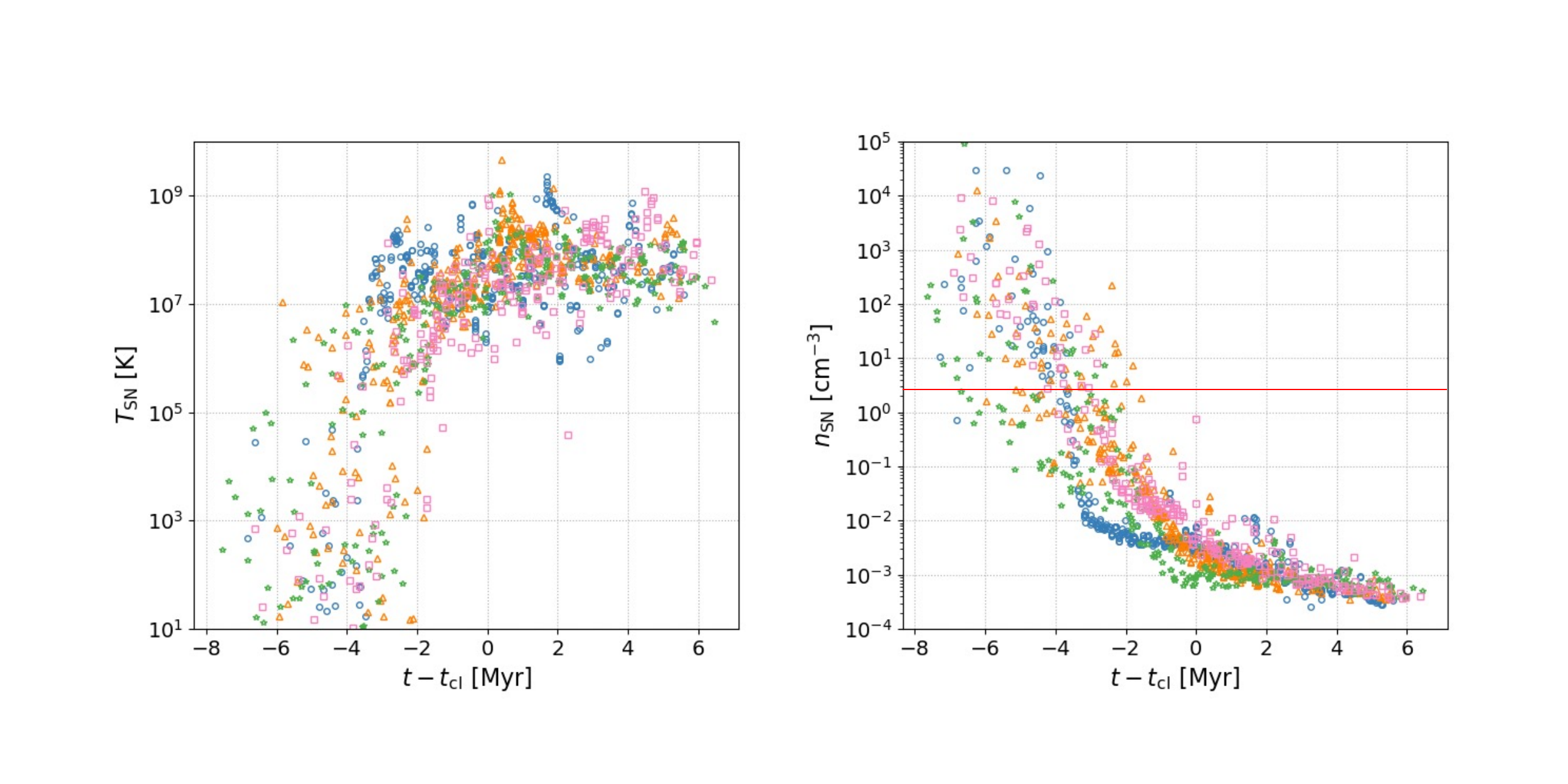}
	\caption{
		SN ambient temperature ($T_{\rm SN}$, left) and density ($n_{\rm SN}$, right) as a function of 
		the normalized time ($t - t_{\rm cl}$) for the top four SN clusters 
		(ranked by $N_{\rm cl}$) 
		in the \textit{gizmo*} model.
		Each point represents an SN event
		and different colors and symbols represent different clusters.
		{The horizontal red line indicates the density above which the SN cooling radius becomes unresolved.}		
		The first SNe occur in cold and dense gas
		which heat up and evacuate the gas,
		and the subsequent SNe occur in increasingly hot and diffuse gas.
	}
	\label{fig:fofntvstime}
\end{figure*}

\begin{figure*}
	\centering
	\includegraphics[trim=1cm 1cm 1cm 1cm,clip, width=0.99\linewidth]{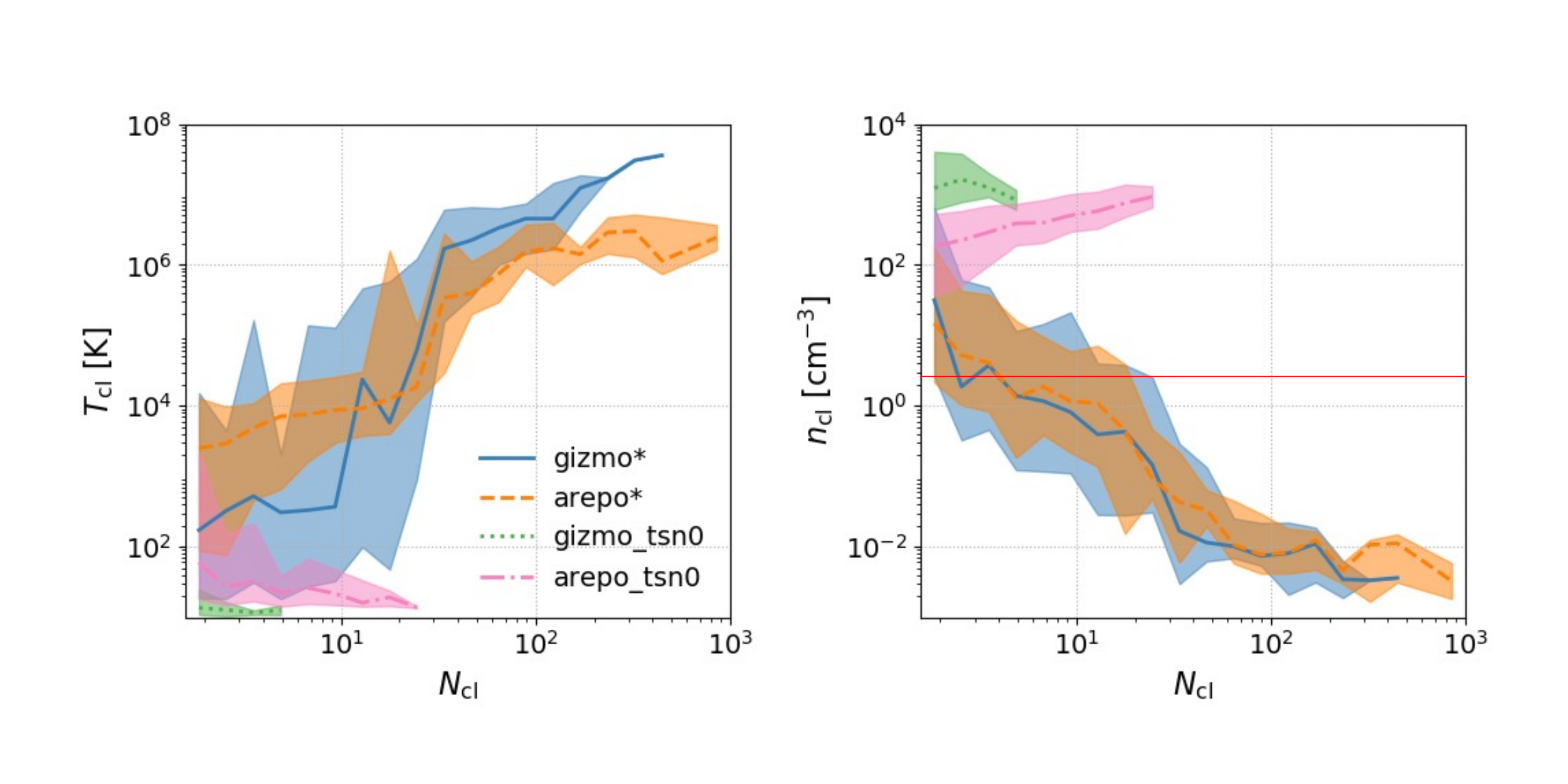}
	\caption{
		\textit{Left:}		
		median SN ambient temperature of a cluster ($T_{\rm cl}$) as a function of $N_{\rm cl}$.
		The solid line shows the median values in each $N_{\rm cl}$ bin while the shaded area brackets the 25\% and 75\% percentiles.
		\textit{Right:}
		same as \textit{left} but for the median SN ambient density $n_{\rm cl}$.
		{The horizontal red line indicates the density above which the SN cooling radius becomes unresolved.}		
		Hot gas ($T > 10^5$~K) is generated when $N_{\rm cl} \gtrsim 20$.
	}
	\label{fig:foftvsnclus}
\end{figure*}

In this section, 
we study the clustering properties of SNe
and show that 
SN clustering 
has a fundamental impact on
$n_{\rm SN}$, SFR burstiness, and gas morphology. 
The effect of SN clustering is two-fold.
Firstly,
it reduces the ambient gas densities 
such that the SN bubble will retain the over-pressurized hot gas for a longer time 
as the radiative cooling rate becomes lower than the energy injection rate 
of the temporally clustered SNe
\citep{2015ApJ...802...99K,2015MNRAS.449.1057G,2015MNRAS.454..238W,2016MNRAS.456.3432G, 2017MNRAS.465.2471G}.
Secondly,
it generates coherent gas flows with less momentum cancellation.
Therefore,
it facilitates the formation of large SN bubbles that can break out of the gaseous disk \citep{2017ApJ...834...25K, 2022ApJ...932...88O}.

The SN clusters are defined by
the friends-of-friends method
following \citet{2021MNRAS.506.3882S}.
This is possible as we record the location and time of each SN event in the simulations.
In addition to a linking length of 10~pc,
we also impose a linking time of 1~Myr
to ensure that the SNe are indeed temporally clustered.
The number of SNe in each SN cluster 
is defined as the clustering number $N_{\rm cl}$.
{We perform the analysis in the ``lab'' frame,
and do not correct for the effect of galactic rotation.
This means that an SN might drift away from the SN cluster it physically belongs 
by more than a linking length due to rotation
and thus it would be incorrectly excluded. 
We argue that this should only be a minor effect as most of our SNe 
are temporally clustered at much shorter timescales than 1~Myr (see Fig.~\ref{fig:foftvsnclus}). 
}

Fig.~\ref{fig:fofscatterall}
shows the spatial distributions of the top ten SN clusters (ranked by $N_{\rm cl}$) in different models.
Each point represents an SN event
and different colors represent different SN clusters.
The average clustering number of the top ten SN clusters $\langle N_{\rm cl} \rangle$ is shown in each panel.

Comparing between the fiducial models,
\textit{gizmo*} and \textit{arepo*} show significantly more clustered SNe than \textit{ramses},
with $\langle N_{\rm cl} \rangle$ larger by about an order of magnitude.
In Lagrangian codes,
as gas is able to collapse to densities far above the threshold density,
star formation is locally enhanced due to the $\rho_{\rm g}^{1.5}$ dependency,
%$\epsilon_{\rm SF}$ is effectively enhanced 
This makes the SNe highly clustered,
which collectively generates large SN bubbles and low $n_{\rm SN}$, and
leads to bursty star formation.
{The significantly less clustered SNe in \textit{ramses} suggests that having low $n_{\rm SN}$,
as we showed in Fig.~\ref{fig:sncdfcompare}, 
is not a gaurantee for efficient feedback -- the SNe have to be clustered too.
}
Meanwhile,
the location of SNe in \textit{gizmo*} and \textit{arepo*}
closely follows the trajectory of galactic rotation,
implying that the SN progenitors rarely drift away their birth clouds
and explode in the diffuse medium,
which is a potential cause for the low $n_{\rm SN}$.
This is perhaps not surprising
as we do not resolve the collisional stellar dynamics 
and neither do we include any sub-grid treatment 
for the so-called ``run-away'' or ``walk-away'' stars \citep{2014ApJ...782....7D}.

{To further support our argument that
strong SN clustering is indeed caused by locally enhanced SFR,
we conduct another numerical experiment,
presented in Appendix~\ref{app:tSFR2Gyr},
where we force the local SFR to scale linearly rather than super-linearly with gas density.
In this case, despite the formation of dense clouds, 
star formation in the clouds proceeds slowly, resulting in low clustering of SNe and an attendant lack of burstiness (resulting in inefficient outflows -- see the following section). 
In this way, the Lagrangian code behaves similarly to the fiducial Ramses model 
despite using the fiducial SN delay time of 10 Myr, 
emphasizing that the root difference is tied ultimately to SN clustering.
%to be $\dot{\rho}_{\rm *} = \rho_{\rm gas} / t_{\rm SFR}$
%with a constant $t_{\rm SFR} = 2$~Gyr.
%As $\dot{\rho}_{\rm *}$ scales linearly rather than super-linearly with density,
%there is no locally enhanced SFR
}

On the other hand,
the instantaneous SN models (\textit{gizmo\_tsn0}, \textit{arepo\_tsn0}, and \textit{ramses\_tsn0}) 
represent the extreme case
where SN clustering is strongly suppressed.
Once star formation occurs,
SN feedback immediately stops any further gravitational collapse
and the development of any subsequent SNe.
This leads to high $n_{\rm SN}$ and thus more rapid energy loss, 
as well as smaller SN-driven bubbles,
resulting in a weaker dynamical impact on the ISM.

With $\epsilon_{\rm SF} = 100\%$,
SNe are highly clustered in all models.
The Lagrangian codes show a similar level of SN clustering 
compared to their fiducial counterparts (where $\epsilon_{\rm SF} = 1\%$).
The lack of clusters in the central area of \textit{arepo\_sfe100}
is caused by its strong initial blowout.
As opposed to the \textit{ramses} model,
\textit{ramses\_sfe100} shows significantly more clustered SNe, 
with $\langle N_{\rm cl} \rangle$ larger by a factor of 30,
consistent with its low $n_{\rm SN}$, bursty SFR, and large SN bubbles.

Fig.~\ref{fig:foftime} shows
$N_{\rm cl}$ as a function of the median time of the SNe in a cluster ($t_{\rm cl}$)
in the \textit{gizmo*} model.
Each circle represents a SN cluster.
The effect of the SN delay time on clustering is clearly demonstrated:
once $t_{\rm SN}$ is switched from 0 to 10 Myr at $t = 250$~Myr
(indicated by the vertical dashed line),
SNe immediately become significantly clustered.

To demonstrate the development of SN bubbles,
Fig.~\ref{fig:fofntvstime}
shows the
SN ambient temperature ($T_{\rm SN}$, left panel) and density (right panel) 
as a function of 
the normalized time ($t - t_{\rm cl}$) for the the top four SN clusters 
(ranked in $N_{\rm cl}$) 
in the \textit{gizmo*} model.
Each point represents a SN event
and different colors and symbols represent different clusters.
All of these clusters share a similar time evolution:
the first SNe occur in cold and dense gas
which gradually heat up and evacuate the gas,
%such that the subsequent SNe occur in increasingly hot and diffuse gas with less efficient radiative cooling,
eventually saturating at 
$T_{\rm SN} \sim 10^8$~K and $n_{\rm SN} \sim 10^{-3}~{\rm cm^{-3}}$.

To quantify
the environmental properties of SN clusters,
we calculate the median of the ambient temperature ($T_{\rm cl}$) 
and density ($n_{\rm cl}$)
of each SN cluster
as a function of $N_{\rm cl}$
in Fig.~\ref{fig:foftvsnclus}.
The solid line shows the median values of $T_{\rm cl}$ or $n_{\rm cl}$ in each $N_{\rm cl}$ bin 
while the shaded area brackets the 25\% and 75\% percentiles.
For \textit{gizmo\_tsn0} and \textit{arepo\_tsn0},
SNe are forced to occur in cold and dense gas by construction.
In contrast,
in both \textit{gizmo} and \textit{arepo},
as $N_{\rm cl}$ increases,
$T_{\rm cl}$ increases while $n_{\rm cl}$ decreases.
At $N_{\rm cl} > 20$,
$T_{\rm cl}$ becomes higher than $10^5$~K,
indicating the development of hot gas in the SN bubbles,
which has been shown to be critical for launching galactic outflows 
\citep{2015MNRAS.454..238W, 2016MNRAS.456.3432G, 2019MNRAS.483.3363H}.

\subsection{Galactic outflows}

\begin{figure}
	\centering
	\includegraphics[width=0.99\linewidth]{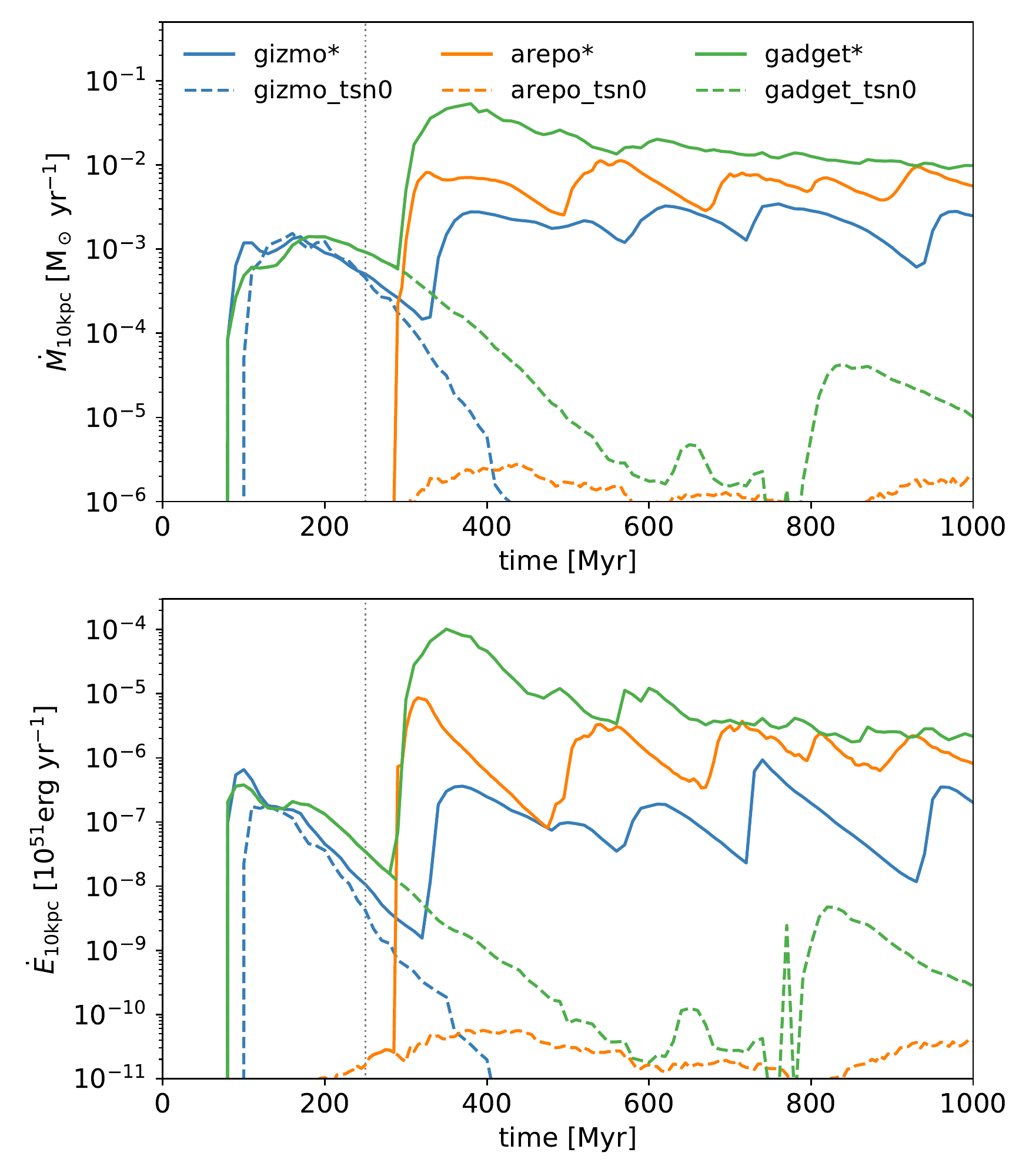}
	\caption{
		Mass (top panel) and energy (bottom panel) outflow rates as a function of time across the planes of $z = \pm 10$ kpc 
		for the Lagrangian models.
		The vertical dotted line indicates $t = 250$~Myr.
		SN delay time has a significant impact on the outflow rates.
	}
	\label{fig:L_OFRvsTime}
\end{figure}

\begin{figure}
	\centering
	\includegraphics[width=0.99\linewidth]{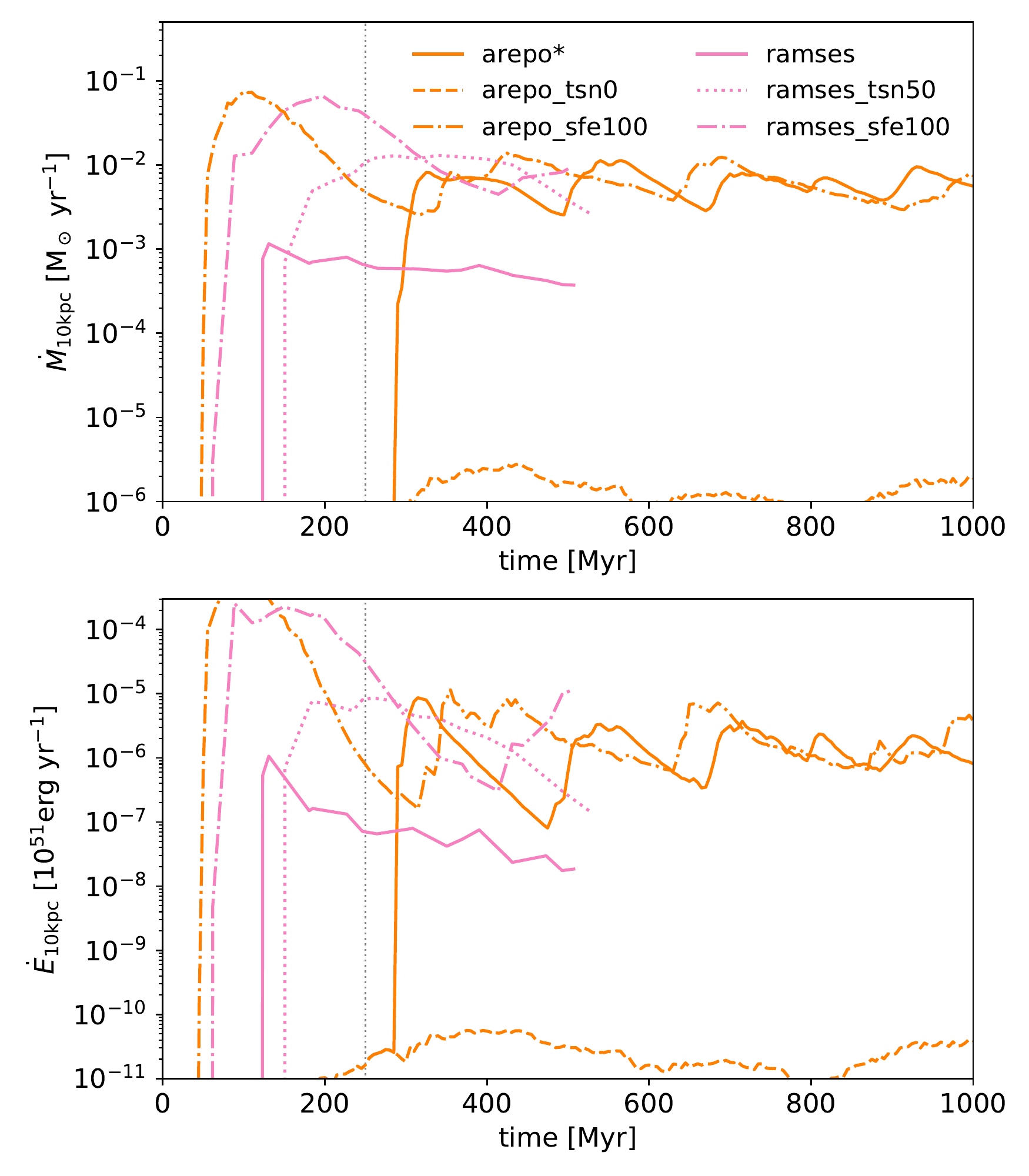}
	\caption{
		Same as Fig.~\ref{fig:L_OFRvsTime} but
		for the {\sc Arepo} and {\sc Ramses} models.
	}
	\label{fig:EL_OFRvsTime}
\end{figure}

%In this section,
%we investigate the 
Galactic outflows driven by SNe play a critical role in galaxy formation.
We characterize the outflows in a similar fashion as in \citet{2019MNRAS.483.3363H}.
We define 
the mass outflow rate as
\begin{equation}
	\dot{M}_{\rm out} = \sum_{i} \frac{m_{\text{g},i} v_{\text{z},i}}{\Delta z}
\end{equation}
and 
the energy outflow rate as
\begin{equation}\label{eq:Eload}
\dot{E}_{\rm out} = \sum_{i} \frac{m_{\text{g},i} (v^2_{i}  +  \gamma u_i) v_{\text{z},i} }{\Delta z},
\end{equation}
where 
$v_i$ is the gas velocity,
$v_{\text{z},i}$ is the gas velocity along the vertical direction,
$\gamma = 5/3$ is the adiabatic index,
$u_i$ is the specific internal energy,
and
$\Delta z = 0.1$~kpc is the thickness of the plane where outflows are measured.
Here $z = 0$ corresponds to the mid-plane of the disk in the initial conditions.
The summation is over cells or particles 
that are
(i) located between $z = 10 \pm 0.5 \Delta z$~kpc or $z = -10 \pm 0.5 \Delta z$~kpc
and 
(ii) traveling ``outward'', i.e., $ z_{i} v_{\text{z},i} > 0$.

We define the
mass loading factor
\begin{equation}
\eta_{\rm m} \equiv \frac{ \dot{M}_{\rm out} }{ \langle\text{SFR}\rangle }
\end{equation}
and the 
energy loading factor
\begin{equation}
\eta_{\rm e}  \equiv  \frac{M_{\rm SN} \dot{E}_{\rm out} }{ E_{\rm SN}  \langle\text{SFR}\rangle }
\end{equation}
where
$E_{\rm SN} = 10^{51}$~erg is the injection energy per SN
and
$M_{\rm SN} = 100~{\rm M_\odot}$ 
is the corresponding mass formed in a stellar population per SN.
The bracket $\langle ... \rangle$ represents time averaging,
and we exclude $t < 250$~Myr 
to discard the initial transient phase.
We use the time-averaged SFR instead of the instantaneous SFR as the normalization factor
such that the fluctuations of $\eta_{\rm m}$ and $\eta_{\rm e}$
are purely due to the fluctuations in the outflows.
The time-averaged 
$\eta_{\rm m}$ and $\eta_{\rm e}$,
summarized in Table~\ref{tab:sum_table},
are therefore
$\langle\eta_{\rm m} \rangle = \langle\dot{M}_{\rm out}\rangle / \langle\text{SFR}\rangle$
and 
$\langle \eta_{\rm e} \rangle =  M_{\rm SN} \langle\dot{E}_{\rm out}\rangle / (E_{\rm SN}  \langle\text{SFR}\rangle  )$, 
respectively.

Fig.~\ref{fig:L_OFRvsTime} shows 
$\dot{M}_{\rm out}$ (top panel) 
and
$\dot{E}_{\rm out}$ (bottom panel) 
as a function of time across the planes of $z = \pm 10$ kpc 
for the Lagrangian models.
Instantaneous SN models have more than two orders of magnitude lower outflow rates compared to the fiducial models,
demonstrating that 
SN clustering substantially enhances outflows.
%as it increases the fraction of SNe
%occuring at low densities where radiative cooling is inefficient.
Both \textit{gadget\_tsn0} and \textit{gizmo\_tsn0} show weak but non-zero outflows
at $t \sim 100$~Myr as a result of the initial star formation,
which is absent in \textit{arepo\_tsn0}.
This is likely due to the difference in the initial conditions:
\textit{arepo\_tsn0} includes a low-density background for numerical purposes
while \textit{gadget\_tsn0} and \textit{gizmo\_tsn0} adopt a vacuum background
which presumably facilitates the relatively weak outflows to expand and fill up the vacuum.
%he outflow rates closely reflect the SN clustering.
%while the \textit{ramses} model have intermediate outflow rates.

Interestingly,
the outflow rates in the fiducial models show notable differences 
even though
their SFRs, $n_{\rm SN}$, and $n_{\rm SF}$ are all in good agreement.
The difference between 
\textit{arepo*} and \textit{gadget*} is mostly due to their SFRs, which differ by a factor of three,
while their 
$\langle\eta_{\rm m} \rangle$
and 
$\langle\eta_{\rm e} \rangle$
only differ by a factor of 25\% and 80\%, respectively.
However,
\textit{gizmo*} 
is
a factor of 4 to 5 lower in
$\langle\eta_{\rm m} \rangle$
and
a factor of 10 to 18 lower in
$\langle\eta_{\rm e} \rangle$
compared to 
\textit{arepo*} and
\textit{gadget*}.
The origin of this difference is still unclear
and requires further investigation in future work.

%(marked as the vertical dotted line),

We now
compare outflows
between Eulerian and Lagrangian codes 
in Fig.~\ref{fig:EL_OFRvsTime}.
For clarity,
we only show the {\sc Arepo} models as a representation of Lagrangian codes.
%as it shows intermediate outflow rates among the three Lagrangian codes.
%\textit{arepo\_sfe100} shows comparable outflow rates with \textit{arepo*}.
The \textit{ramses} model 
is
a factor of 32 lower in
$\langle\eta_{\rm m} \rangle$
and
a factor of 100 lower in
$\langle\eta_{\rm e} \rangle$
compared to 
\textit{arepo*},
reflecting the fact that SNe are significantly more clustered in \textit{arepo*}.
%shows a significantly lower outflow rates
%compared to \textit{arepo*} due to the difference in SN clustering.
The outflow rates in {\sc Ramses}
are enhanced
with either a larger $t_{\rm SN}$ or 
a larger $\epsilon_{\rm SF}$,
as both would increase the SN clustering.
Indeed,
\textit{ramses\_tsn50} and
\textit{ramses\_sfe100}
both show comparable outflow rates with \textit{arepo*}.

We conclude that SN clustering plays a fundamental role in driving of galactic outflows.
%The difference between Eulerian and Lagrangian codes 
%in outflow rates 
%stems from their difference in SN clustering.
Only models with significant SN clustering show $\langle\eta_{\rm m} \rangle$ larger than unity.

\begin{deluxetable*}{l c c c c c}
	%\tablenum{1}
	\tablecaption{
		Summary of simulation results. 
	}
	\label{tab:sum_table}
	\tablewidth{0pt}
	\tablehead{
		\colhead{model name} &
		\colhead{initial blowout} &		
		\colhead{clustered SNe} &
		\colhead{$\langle\text{SFR}\rangle$} &		
		\colhead{$\langle \eta_{\rm m} \rangle$} &
		\colhead{$\langle\eta_{\rm e} \rangle$} 
		%%%%%%%%%%%%%%%%%%%%%%%
	}
	\decimalcolnumbers
	\startdata
gizmo            &     yes  &     yes    &   1.2  		 &  3.0    &	0.041   \\
arepo            &     yes  &     yes    &   0.39 	   &  19    &	1.1   \\
ramses           &     no   &     no     &   3.1       &  0.17   & 0.0015   \\
gizmo\_tsn0      &     no   &     no     &   1.8    	 &  0.011  &	$5.6\times 10^{-5}$   \\
arepo\_tsn0      &     no   &     no     &   0.82 	   &  0.0017 &	$3.3\times 10^{-6}$   \\
ramses\_tsn0     &     no   &     no     &   2.2       &  -       & -   \\
gadget\_tsn0     &     no   &     no     &   1.7 		   &  0.032  &	$8.1\times 10^{-5}$   \\
gizmo*           &     no   &     yes    &   3.1 		   &  1.1    &	0.015   \\
arepo*           &     no   &     yes    &   1.1 		   &  5.4    &	0.15   \\
gadget*          &     no   &     yes    &   3.4 	     &  4.3    &	0.27   \\
gizmo\_sfe100    &     yes  &     yes    &   0.66 		 &  16    &	0.25   \\
arepo\_sfe100    &     yes  &     yes    &   0.85 	   &  7.6    &	0.29   \\
ramses\_sfe100   &     yes  &     yes    &   0.53 	   &  13    &	0.49   \\
ramses\_tsn50    &     yes  &     yes    &   3.1       &  3.2    & 0.10   \\	
	\enddata
	\tablecomments{	
		(4) time-averaged SFR in units of $10^{-3}~{\rm M_\odot~yr^{-1}}$.
		(5) time-averaged mass loading factor.
		(6) time-averaged energy loading factor.
		The first 250~Myr is excluded from the time averaging to discard the initial transient phase.
	}
\end{deluxetable*}

\section{Discussion} \label{sec:diss}

\subsection{Comparison with previous work}

Several recent studies have quantified
outflows in terms of $\eta_{\rm m}$ and $\eta_{\rm e}$ 
in similar initial conditions of a dwarf galaxy 
with resolved SN feedback.
%\citep{2019MNRAS.483.3363H, 2021MNRAS.506.3882S, 2021MNRAS.501.5597G}.
%For comparison purposes,
%we focus on models with SN feedback but without photoionization.
Our fiducial Lagrangian models
(\textit{gizmo*},
\textit{arepo*},
and \textit{gadget*})
are in broad agreement with these studies:
\citet{2019MNRAS.483.3363H} 
found 
$\langle \eta_{\rm m} \rangle \sim 4$
and 
$\langle \eta_{\rm e} \rangle \sim 0.07$
at $|z| = 10$~kpc
using the {\sc Gadget} code.
%Though the time-averaged values were not reported,
\citet{2021MNRAS.506.3882S}
found 
$\eta_{\rm m}$ fluctuating between 1 and 10
and 
$\eta_{\rm e}$ between 0.003 -- 0.3
at $|z| = 10$~kpc
in a comparable model (their \textit{PE-SN})
using the {\sc Arepo} code.
\citet{2021MNRAS.501.5597G}
found 
$\eta_{\rm m}$ between 5 and 10
and 
$\eta_{\rm e}$ between $10^{-4}$ and $10^{-3}$
at $|z| = 2$~kpc
in a comparable model (their {\sc Fixed\_SN\_energy})
using the {\sc Arepo} code.
We note that $\eta_{\rm m}$ is in a better agreement among these studies
than $\eta_{\rm e}$,
probably because $\eta_{\rm e}$ is very sensitive to
small differences in the gas velocity (see Eq. \ref{eq:Eload}).
Our results confirm these previous findings that
the predicted $\eta_{\rm m}$ from SN-resolved galaxy scale simulations are  significantly lower than what cosmological simulations commonly assume in their sub-grid models.
Observations of nearby dwarf galaxies seem to support the low $\eta_{\rm m}$ case \citep{2019ApJ...886...74M}.

In a similar setup,
\citet{2017MNRAS.471.2151H} 
found that
a shortened SN delay time ($t_{\rm SN} = 3$~Myr)
has the effect of suppressing the formation of SN bubbles and galactic outflows,
which is consistent with our results.
However, they did not find an effect of $t_{\rm SN}$ on the burstiness of star formation.
\citet{2021MNRAS.506.3882S}
showed that photoionization from massive stars prior to the SN events (the so-called ``early feedback'')
reduces SN clustering
and thus significantly suppresses the burstiness of star formation, formation of SN bubbles, and galactic outflows,
reducing $\eta_{\rm m}$ and $\eta_{\rm e}$ by more than an order of magnitude.
This is very similar to our instantaneous SN models,
which can be viewed as an extreme case of early feedback.
\citet{2022MNRAS.512..199K} simulated a Milky Way-like galaxy with $m_{\rm g} \sim 10^5~{\rm M_\odot}$
where SN feedback is modeled in a sub-grid fashion.
Although the simulated galaxy and the numerical treatments are very different,
they found that a longer SN delay time enhances galactic outflows
by enhancing the clustering of young stars,
which is qualitatively consistent with our results.

\citet{2016ApJ...833..202K}
have conducted a detailed code comparison study 
for an isolated Milky Way-like galaxy with significantly more participating codes.
They found that different codes generally agree well between each other.
In particular,
they did not find systematically differences between
Lagrangian and Eulerian codes
in terms of burstiness and the sizes of SN bubbles.
This is probably because
they have adopted the simple thermal injection
as their sub-grid prescription for SN feedback,
which is known to be very inefficient at this resolution
as most energy would be radiated away without generating much momentum.

\subsection{Differences between numerical methods}

As we have shown in Section \ref{sec:nSFnSN},
the differences between Lagrangian and Eulerian codes
arise near the resolution limit when the Jeans length becomes unresolved.
This is perhaps not surprising:
code differences almost by definition have to occur near the resolution limit,
as the resolved scales should converge to the physical solutions for any numerically consistent method.
However,
in our case,
the differences at the resolution limit
quickly propagate to much larger, well-resolved scales 
due to clustered star formation and SN feedback.

As we alluded to in Section \ref{sec:nSFnSN},
once the gas enters the Jeans length-unresolved regime,
both methods can no longer faithfully follow the collapse
and the evolution of gas depends sensitively on numerics 
such as gravitational softening.
That said,
it is interesting to ask which method is ``less wrong''.
For a collapsing cloud that becomes unresolved,
both the Jeans mass and Jeans length would keep decreasing as the density increases.
In other words,
the cloud should keep collapsing and fragmenting.
Lagrangian codes cannot capture 
the fragmentation without particle splitting or cell refinement,
and they might underestimate the collapsing speed if the adopted softening length is much larger than the cell size.
However, the cloud will continue to collapse as expected.
In contrast,
the collapse in an Eulerian code would halt when the Jeans length becomes unresolved
as it requires accreting gas into the minimal cell.
In this sense,
Lagrangian codes are arguably more accurate than Eulerian codes in the unresolved regime\footnote{
Eulerian codes might behave more similarly to Lagrangian codes
when adopting a minimal cell size well below the resolvable Jeans length.
However, this would be prohibitively expensive in practical applications.}.

The situation becomes more subtle if we consider the physical processes we do not include in this work.
In particular,
as we do not include feedback from photoionization,
we might have overestimated SN clustering and more significantly so in our Lagrangian models. 
If this is the case,
our Eulerian model would more accurately, though coincidentally, predict the SN clustering.

%In order to resolve individual SNe,
%resolutions better than our high-resolution models
%(12.5 for Lagrangian codes and 3.5 for Eulerian codes) are required.

%Our results demonstrate the importance of SN clustering
%as a key factor controlling the burstiness and galactic outflows.

\subsection{Star formation efficiency}

Previous studies have shown that 
the star formation efficiency
only has a weak effect on the galaxy scale SFR 
in both Lagrangian and Eulerian codes
(e.g., \citealp{2013ApJ...770...25A, 2016MNRAS.462.3053B, 2017ApJ...845..133S, 2018MNRAS.480..800H}).
While this is consistent with our Lagrangian models,
it appears to be in conflict with our Eulerian models
where increasing $\epsilon_{\rm SF}$ 
leads to a qualitative change in
the burstiness of star formation due to enhanced SN clustering.
This is probably because
SN feedback is unresolved and treated in a sub-grid fashion in those studies,
which reduces its dynamical impact.
Alternatively,
as those studies simulated more massive galaxies,
the local burstiness might have been averaged out such that the
global SFR remains the same.
In this case,
we expect the outflow rates to increase with $\epsilon_{\rm SF}$.

Instead of assuming a constant $\epsilon_{\rm SF}$,
some recent simulations of galaxy formation have adopted a class of sub-grid models for star formation
that calculates $\epsilon_{\rm SF}$ based on local gas properties 
\citep{2016ApJ...826..200S,2020MNRAS.492.1385K}.
These models assume a sub-grid log-normal density distribution as predicted by
simulations of supersonic turbulence,
and models differ in their criteria for the onset of star formation 
\citep{2005ApJ...630..250K, 2008ApJ...684..395H, 2011ApJ...730...40P, 2012ApJ...761..156F, 2018ApJ...863..118B, 2019ApJ...879..129B}.
Our results suggest that
adopting such a sub-grid model would have 
a much stronger effect in Eulerian codes
as they are more sensitive to the variation of $\epsilon_{\rm SF}$.

%instead of assuming a constant $\epsilon_{\rm SF}$.
%log-normal density distribution 

\subsection{Numerical convergence}

The significant differences between our low- and high-resolution models
suggest that we have not yet reached numerical convergence.
This is consistent with 
\citet{2018MNRAS.478..302S}
where their models with
$m_{\rm g} = 20~{\rm M_\odot}$ and
$m_{\rm g} = 200~{\rm M_\odot}$
still show significantly different outflow rates.
In addition,
the convergence study in
\citet{2019MNRAS.483.3363H} 
with $m_{\rm g} = 1, 5, 25$, and $125~{\rm M_\odot}$
showed that convergence was only achieved at $m_{\rm g} = 5~{\rm M_\odot}$
when individual SNe are properly resolved.
Therefore,
we optimistically expect that our high-resolution models are actually close to convergence.

Many recent cosmological ``zoom-in'' simulations
have adopted
a SN feedback model
that injects thermal energy for resolved SNe
and injects the terminal momentum for unresolved SNe as a sub-grid model
(e.g., \citealp{2015ApJ...804...18A, 2014MNRAS.445..581H, 2018MNRAS.480..800H, 2015MNRAS.451.2900K, 2019MNRAS.489.4233M}).
Most of these simulations
have resolutions much coarser than our low-resolution models,
suggesting that their SN feedback still operates in the sub-grid regime
where galactic outflows are expected to be underestimated
due to the lack of hot gas \citep{2019MNRAS.483.3363H}. 
The more recent studies such as
\citet{2020MNRAS.491.1656A, 2022MNRAS.513.1372G, 2022MNRAS.516.5914C}
have started to resolve individual SNe
in cosmological ``zoom-in'' simulations of dwarf galaxies,
which is a promising way forward.

\subsection{Missing physics and future prospects}

The main caveat of our results is the fact that
we do not include pre-SN feedback such as photoionization.
As we already discussed,
this could lead to overestimated SN clustering,
in particular for our Lagrangian models.
This is a potential solution for the discrepancies we find 
between the two types of methods.
An interesting follow-up project
is therefore to include photoionization
and see if the discrepancies can be mitigated.
On the other hand,
the suppression of SN clustering by photoionization
in \citet{2021MNRAS.506.3882S}
could also have been overestimated 
by the adopted ``Str\"{o}mgren-type'' approach 
where dense clumps are preferentially ionized.
{Interestingly,
\citet{2021MNRAS.504.1039R} conducted ISM-patch simulations 
using radiative transfer based on an inverse ray-tracing techique
and reached the same conclusion
that photoionization suppresses SN clustering which leads to less efficient hot gas generation.
}
This should be investigated
by future galaxy scale simulations
using more accurate methods of radiative transfer
such as adaptive ray tracing \citep{2019MNRAS.482.1304E}.
We note that including a solver for radiative transfer 
does not necessarily resolve the issue, 
as the important part is to ensure sufficient angular resolution 
to prevent the dense clumps from being preferentially ionized.
This therefore implies a very demanding computational cost.

Another potentially important element is the collisional stellar dynamics.
While we adopt a collisionless gravity solver in this work,
the stellar dynamics in young star clusters is actually collisional.
As our results highlight the importance of SN clustering,
it is desirable to 
include an accurate N-body integrator 
such as 
\citet{2020ApJ...904..192W} and \citet{2021MNRAS.502.5546R}
to properly follows the evolution of young star clusters.
However,
the computational feasibility of incorporating it into galaxy scale simulations
remains to be explored.
On the other hand, 
stellar dynamics can be modeled in a sub-grid fashion.
\citet{2022arXiv220509774S}
simulated a dwarf galaxy similar to ours
including a sub-grid model for runaway massive stars
which can enhance
$\langle \eta_{\rm m} \rangle$
by 50\%
and 
$\langle \eta_{\rm e} \rangle$
by a factor of five.

\section{Summary} \label{sec:sum}

We have conducted a suite of simulations of an isolated dwarf galaxy 
using four different hydrodynamical codes 
({\sc Gizmo}, {\sc Arepo}, {\sc Gadget}, and {\sc Ramses}).
All codes adopt the same physical model,
which includes radiative cooling, photoelectric heating, star formation, and SN feedback.
We directly resolve individual SN feedback without using sub-grid models,
which is a major source of uncertainty in cosmological simulations.
Our main results can be summarized as follows.

\begin{itemize}

\item 	
The time-averaged SFRs and the distributions of gas density and temperature 
are in reasonable agreement in all codes (Table~\ref{tab:sum_table} and Fig.~\ref{fig:pdcodecompare}).
However, 
Lagrangian codes show a burstier star formation history (Figs.~\ref{fig:sfrvstimefid} and \ref{fig:sfrvstimetsn0}),
larger SN-driven bubbles (Fig.~\ref{fig:maps_all_scale}),
and stronger galactic outflows (Figs.~\ref{fig:L_OFRvsTime} and \ref{fig:EL_OFRvsTime}),
in striking contrast to the Eulerian code.
This originates from the different behaviors 
as gas collapses beyond the star formation threshold:
the Jeans-length unresolved gas
collapses to much higher densities in the Lagrangian codes (Fig.~\ref{fig:sfcdf}),
leading to more complete conversation of gas into stars and hence more
highly clustered SNe (Fig.~\ref{fig:fofscatterall}). 
Hot gas ($T > 10^5$~K) in the SN bubbles that drives galactic outflows is generated 
when the SN clustering number is sufficiently high (Fig.~\ref{fig:foftvsnclus}).

%In Lagrangian codes,
%this unresolved gas keeps collapsing to densities much higher than the threshold density.
%In Eulerian codes,
%the unresolved gas collapses much slower due to the fixed minimal cell size
%and it remains at the threshold density.

%\item 
%The dynamical impact of SN feedback is substantially enhanced by clustering,
%as the first SNe reduce the ambient gas densities 
%and thus the subsequent SNe are subject to less energy loss from radiative cooling (Figs.~\ref{fig:sncdfcompare} and \ref{fig:fofntvstime}).

\item 
If we let SN feedback occur with a zero delay time 
%($t_{\rm SN} = 0$) 
immediately after star formation as a numerical experiment,
SN clustering would be strongly suppressed and SNe are forced to occur at high densities with rapid radiative losses.
In this case, all codes behave similarly,
showing a non-fluctuating SFR,
no visible SN bubbles,
and vanishing galactic outflows.

\item 
The adopted star formation efficiency ($\epsilon_{\rm SF}$)
has a significant effect on
SN clustering in Eulerian codes,
which in turn affects
the star formation burstiness, sizes of SN bubbles, and outflow rates.
%become comparable to the Lagrangian codes.
In contrast,
$\epsilon_{\rm SF}$
only plays a minor role
in Lagrangian codes
where gas collapses to much higher densities
such that local star formation is significantly enhanced,
effectively enhancing $\epsilon_{\rm SF}$ even when a low value is used.

\item 
Lagrangian models are in good agreement with each other in terms of gas morphology, SN densities, and star formation densities.
However, {\sc Gizmo} shows notably weaker outflows compared to {\sc Arepo} and {\sc Gadget} in the fiducial models,
which requires further investigations in future work.

%($\epsilon_{\rm SF} = 100\%$)

\end{itemize}

\section*{Acknowledgments}
{We thank the anonymous referee for the constructive comments that helped improve the manuscript,
in particular for suggesting the additional run in Appendix~\ref{app:tSFR2Gyr}.}
We thank Volker Springel and Chang-Goo Kim
for stimulating discussion.
C.Y.H. acknowledges support from the Deutsche Forschungsgemeinschaft (DFG, German Research Foundation) via German-Israel Project Cooperation grant STE1869/2-1 GE625/17-1
and NASA ATP grant 80NSSC22K0716.
The work of MCS was supported by a grant from the Simons Foundation (CCA 668771, LEH) and by the DFG under Germany’s Excellence Strategy EXC 2181/1-390900948 (the Heidelberg STRUCTURES Excellence Cluster).
The Center for Computational Astrophysics is supported by the Simons Foundation.
GLB acknowledges support from the NSF (AST-2108470, XSEDE), a NASA TCAN award, 
and the Simons Foundation through their support of the Learning the Universe Collaboration.
B.B. thanks the Alfred P. Sloan Foundation and the Packard Foundation for support as well as the support of NASA grant No. 80NSSC20K0500.
Y.L. acknowledges financial support from NSF grant AST- 2107735 and NASA grant 80NSSC22K0668.

\bibliography{literatur}{}

\begin{thebibliography}{}
\expandafter\ifx\csname natexlab\endcsname\relax\def\natexlab#1{#1}\fi
\providecommand{\url}[1]{\href{#1}{#1}}
\providecommand{\dodoi}[1]{doi:~\href{http://doi.org/#1}{\nolinkurl{#1}}}
\providecommand{\doeprint}[1]{\href{http://ascl.net/#1}{\nolinkurl{http://ascl.net/#1}}}
\providecommand{\doarXiv}[1]{\href{https://arxiv.org/abs/#1}{\nolinkurl{https://arxiv.org/abs/#1}}}

\bibitem[{{Agertz} \& {Kravtsov}(2015)}]{2015ApJ...804...18A}
{Agertz}, O., \& {Kravtsov}, A.~V. 2015, \apj, 804, 18,
  \dodoi{10.1088/0004-637X/804/1/18}

\bibitem[{{Agertz} {et~al.}(2013){Agertz}, {Kravtsov}, {Leitner}, \&
  {Gnedin}}]{2013ApJ...770...25A}
{Agertz}, O., {Kravtsov}, A.~V., {Leitner}, S.~N., \& {Gnedin}, N.~Y. 2013,
  \apj, 770, 25, \dodoi{10.1088/0004-637X/770/1/25}

\bibitem[{{Agertz} {et~al.}(2020){Agertz}, {Pontzen}, {Read}, {Rey}, {Orkney},
  {Rosdahl}, {Teyssier}, {Verbeke}, {Kretschmer}, \&
  {Nickerson}}]{2020MNRAS.491.1656A}
{Agertz}, O., {Pontzen}, A., {Read}, J.~I., {et~al.} 2020, \mnras, 491, 1656,
  \dodoi{10.1093/mnras/stz3053}

\bibitem[{{Bakes} \& {Tielens}(1994)}]{1994ApJ...427..822B}
{Bakes}, E.~L.~O., \& {Tielens}, A.~G.~G.~M. 1994, \apj, 427, 822,
  \dodoi{10.1086/174188}

\bibitem[{{Barnes} \& {Hut}(1986)}]{1986Natur.324..446B}
{Barnes}, J., \& {Hut}, P. 1986, \nat, 324, 446, \dodoi{10.1038/324446a0}

\bibitem[{{Bate} \& {Burkert}(1997)}]{1997MNRAS.288.1060B}
{Bate}, M.~R., \& {Burkert}, A. 1997, \mnras, 288, 1060

\bibitem[{{Benincasa} {et~al.}(2016){Benincasa}, {Wadsley}, {Couchman}, \&
  {Keller}}]{2016MNRAS.462.3053B}
{Benincasa}, S.~M., {Wadsley}, J., {Couchman}, H.~M.~P., \& {Keller}, B.~W.
  2016, \mnras, 462, 3053, \dodoi{10.1093/mnras/stw1741}

\bibitem[{{Bergin} {et~al.}(2004){Bergin}, {Hartmann}, {Raymond}, \&
  {Ballesteros-Paredes}}]{2004ApJ...612..921B}
{Bergin}, E.~A., {Hartmann}, L.~W., {Raymond}, J.~C., \& {Ballesteros-Paredes},
  J. 2004, \apj, 612, 921, \dodoi{10.1086/422578}

\bibitem[{{Burkhart}(2018)}]{2018ApJ...863..118B}
{Burkhart}, B. 2018, \apj, 863, 118, \dodoi{10.3847/1538-4357/aad002}

\bibitem[{{Burkhart} \& {Mocz}(2019)}]{2019ApJ...879..129B}
{Burkhart}, B., \& {Mocz}, P. 2019, \apj, 879, 129,
  \dodoi{10.3847/1538-4357/ab25ed}

\bibitem[{{Calura} {et~al.}(2022){Calura}, {Lupi}, {Rosdahl}, {Vanzella},
  {Meneghetti}, {Rosati}, {Vesperini}, {Lacchin}, {Pascale}, \&
  {Gilli}}]{2022MNRAS.516.5914C}
{Calura}, F., {Lupi}, A., {Rosdahl}, J., {et~al.} 2022, \mnras, 516, 5914,
  \dodoi{10.1093/mnras/stac2387}

\bibitem[{{Cullen} \& {Dehnen}(2010)}]{2010MNRAS.408..669C}
{Cullen}, L., \& {Dehnen}, W. 2010, \mnras, 408, 669,
  \dodoi{10.1111/j.1365-2966.2010.17158.x}

\bibitem[{{de Mink} {et~al.}(2014){de Mink}, {Sana}, {Langer}, {Izzard}, \&
  {Schneider}}]{2014ApJ...782....7D}
{de Mink}, S.~E., {Sana}, H., {Langer}, N., {Izzard}, R.~G., \& {Schneider},
  F.~R.~N. 2014, \apj, 782, 7, \dodoi{10.1088/0004-637X/782/1/7}

\bibitem[{{Dehnen} \& {Aly}(2012)}]{2012MNRAS.425.1068D}
{Dehnen}, W., \& {Aly}, H. 2012, \mnras, 425, 1068,
  \dodoi{10.1111/j.1365-2966.2012.21439.x}

\bibitem[{{Durier} \& {Dalla Vecchia}(2012)}]{2012MNRAS.419..465D}
{Durier}, F., \& {Dalla Vecchia}, C. 2012, \mnras, 419, 465,
  \dodoi{10.1111/j.1365-2966.2011.19712.x}

\bibitem[{{Emerick} {et~al.}(2019){Emerick}, {Bryan}, \& {Mac
  Low}}]{2019MNRAS.482.1304E}
{Emerick}, A., {Bryan}, G.~L., \& {Mac Low}, M.-M. 2019, \mnras, 482, 1304,
  \dodoi{10.1093/mnras/sty2689}

\bibitem[{{Federrath} \& {Klessen}(2012)}]{2012ApJ...761..156F}
{Federrath}, C., \& {Klessen}, R.~S. 2012, \apj, 761, 156,
  \dodoi{10.1088/0004-637X/761/2/156}

\bibitem[{{Forbes} {et~al.}(2016){Forbes}, {Krumholz}, {Goldbaum}, \&
  {Dekel}}]{2016Natur.535..523F}
{Forbes}, J.~C., {Krumholz}, M.~R., {Goldbaum}, N.~J., \& {Dekel}, A. 2016,
  \nat, 535, 523, \dodoi{10.1038/nature18292}

\bibitem[{{Gaburov} \& {Nitadori}(2011)}]{2011MNRAS.414..129G}
{Gaburov}, E., \& {Nitadori}, K. 2011, \mnras, 414, 129,
  \dodoi{10.1111/j.1365-2966.2011.18313.x}

\bibitem[{{Gatto} {et~al.}(2015){Gatto}, {Walch}, {Low}, {Naab}, {Girichidis},
  {Glover}, {W{\"u}nsch}, {Klessen}, {Clark}, {Baczynski}, {Peters},
  {Ostriker}, {Ib{\'a}{\~n}ez-Mej{\'{\i}}a}, \& {Haid}}]{2015MNRAS.449.1057G}
{Gatto}, A., {Walch}, S., {Low}, M.-M.~M., {et~al.} 2015, \mnras, 449, 1057,
  \dodoi{10.1093/mnras/stv324}

\bibitem[{{Gatto} {et~al.}(2017){Gatto}, {Walch}, {Naab}, {Girichidis},
  {W{\"u}nsch}, {Glover}, {Klessen}, {Clark}, {Peters}, {Derigs}, {Baczynski},
  \& {Puls}}]{2017MNRAS.466.1903G}
{Gatto}, A., {Walch}, S., {Naab}, T., {et~al.} 2017, \mnras, 466, 1903,
  \dodoi{10.1093/mnras/stw3209}

\bibitem[{{Gentry} {et~al.}(2017){Gentry}, {Krumholz}, {Dekel}, \&
  {Madau}}]{2017MNRAS.465.2471G}
{Gentry}, E.~S., {Krumholz}, M.~R., {Dekel}, A., \& {Madau}, P. 2017, \mnras,
  465, 2471, \dodoi{10.1093/mnras/stw2746}

\bibitem[{{Girichidis} {et~al.}(2016){Girichidis}, {Walch}, {Naab}, {Gatto},
  {W{\"u}nsch}, {Glover}, {Klessen}, {Clark}, {Peters}, {Derigs}, \&
  {Baczynski}}]{2016MNRAS.456.3432G}
{Girichidis}, P., {Walch}, S., {Naab}, T., {et~al.} 2016, \mnras, 456, 3432,
  \dodoi{10.1093/mnras/stv2742}

\bibitem[{{Glover} \& {Clark}(2012)}]{2012MNRAS.421..116G}
{Glover}, S.~C.~O., \& {Clark}, P.~C. 2012, \mnras, 421, 116,
  \dodoi{10.1111/j.1365-2966.2011.20260.x}

\bibitem[{{Glover} \& {Mac Low}(2007)}]{2007ApJS..169..239G}
{Glover}, S.~C.~O., \& {Mac Low}, M.-M. 2007, \apjs, 169, 239,
  \dodoi{10.1086/512238}

\bibitem[{{Guillet} \& {Teyssier}(2011)}]{2011JCoPh.230.4756G}
{Guillet}, T., \& {Teyssier}, R. 2011, Journal of Computational Physics, 230,
  4756, \dodoi{10.1016/j.jcp.2011.02.044}

\bibitem[{{Gutcke} {et~al.}(2021){Gutcke}, {Pakmor}, {Naab}, \&
  {Springel}}]{2021MNRAS.501.5597G}
{Gutcke}, T.~A., {Pakmor}, R., {Naab}, T., \& {Springel}, V. 2021, \mnras, 501,
  5597, \dodoi{10.1093/mnras/staa3875}

\bibitem[{{Gutcke} {et~al.}(2022){Gutcke}, {Pakmor}, {Naab}, \&
  {Springel}}]{2022MNRAS.513.1372G}
---. 2022, \mnras, 513, 1372, \dodoi{10.1093/mnras/stac867}

\bibitem[{{Habing}(1968)}]{1968BAN....19..421H}
{Habing}, H.~J. 1968, \bain, 19, 421

\bibitem[{{Hennebelle} \& {Chabrier}(2008)}]{2008ApJ...684..395H}
{Hennebelle}, P., \& {Chabrier}, G. 2008, \apj, 684, 395,
  \dodoi{10.1086/589916}

\bibitem[{{Hislop} {et~al.}(2022){Hislop}, {Naab}, {Steinwandel}, {Lah{\'e}n},
  {Irodotou}, {Johansson}, \& {Walch}}]{2022MNRAS.509.5938H}
{Hislop}, J.~M., {Naab}, T., {Steinwandel}, U.~P., {et~al.} 2022, \mnras, 509,
  5938, \dodoi{10.1093/mnras/stab3347}

\bibitem[{{Hopkins}(2015)}]{2015MNRAS.450...53H}
{Hopkins}, P.~F. 2015, \mnras, 450, 53, \dodoi{10.1093/mnras/stv195}

\bibitem[{{Hopkins} {et~al.}(2014){Hopkins}, {Kere{\v s}}, {O{\~n}orbe},
  {Faucher-Gigu{\`e}re}, {Quataert}, {Murray}, \&
  {Bullock}}]{2014MNRAS.445..581H}
{Hopkins}, P.~F., {Kere{\v s}}, D., {O{\~n}orbe}, J., {et~al.} 2014, \mnras,
  445, 581, \dodoi{10.1093/mnras/stu1738}

\bibitem[{{Hopkins} {et~al.}(2018){Hopkins}, {Wetzel}, {Kere{\v s}},
  {Faucher-Gigu{\`e}re}, {Quataert}, {Boylan-Kolchin}, {Murray}, {Hayward},
  {Garrison-Kimmel}, {Hummels}, {Feldmann}, {Torrey}, {Ma},
  {Angl{\'e}s-Alc{\'a}zar}, {Su}, {Orr}, {Schmitz}, {Escala}, {Sanderson},
  {Grudi{\'c}}, {Hafen}, {Kim}, {Fitts}, {Bullock}, {Wheeler}, {Chan},
  {Elbert}, \& {Narayanan}}]{2018MNRAS.480..800H}
{Hopkins}, P.~F., {Wetzel}, A., {Kere{\v s}}, D., {et~al.} 2018, \mnras, 480,
  800, \dodoi{10.1093/mnras/sty1690}

\bibitem[{{Hu}(2019)}]{2019MNRAS.483.3363H}
{Hu}, C.-Y. 2019, \mnras, 483, 3363, \dodoi{10.1093/mnras/sty3252}

\bibitem[{{Hu} {et~al.}(2017){Hu}, {Naab}, {Glover}, {Walch}, \&
  {Clark}}]{2017MNRAS.471.2151H}
{Hu}, C.-Y., {Naab}, T., {Glover}, S.~C.~O., {Walch}, S., \& {Clark}, P.~C.
  2017, \mnras, 471, 2151, \dodoi{10.1093/mnras/stx1773}

\bibitem[{{Hu} {et~al.}(2016){Hu}, {Naab}, {Walch}, {Glover}, \&
  {Clark}}]{2016MNRAS.458.3528H}
{Hu}, C.-Y., {Naab}, T., {Walch}, S., {Glover}, S.~C.~O., \& {Clark}, P.~C.
  2016, \mnras, 458, 3528, \dodoi{10.1093/mnras/stw544}

\bibitem[{{Hu} {et~al.}(2014){Hu}, {Naab}, {Walch}, {Moster}, \&
  {Oser}}]{2014MNRAS.443.1173H}
{Hu}, C.-Y., {Naab}, T., {Walch}, S., {Moster}, B.~P., \& {Oser}, L. 2014,
  \mnras, 443, 1173, \dodoi{10.1093/mnras/stu1187}

\bibitem[{{Hu} {et~al.}(2021){Hu}, {Sternberg}, \& {van
  Dishoeck}}]{2021ApJ...920...44H}
{Hu}, C.-Y., {Sternberg}, A., \& {van Dishoeck}, E.~F. 2021, \apj, 920, 44,
  \dodoi{10.3847/1538-4357/ac0dbd}

\bibitem[{{Hunter} {et~al.}(2010){Hunter}, {Elmegreen}, \&
  {Ludka}}]{2010AJ....139..447H}
{Hunter}, D.~A., {Elmegreen}, B.~G., \& {Ludka}, B.~C. 2010, \aj, 139, 447,
  \dodoi{10.1088/0004-6256/139/2/447}

\bibitem[{{Ib{\'a}{\~n}ez-Mej{\'\i}a}
  {et~al.}(2016){Ib{\'a}{\~n}ez-Mej{\'\i}a}, {Mac Low}, {Klessen}, \&
  {Baczynski}}]{2016ApJ...824...41I}
{Ib{\'a}{\~n}ez-Mej{\'\i}a}, J.~C., {Mac Low}, M.-M., {Klessen}, R.~S., \&
  {Baczynski}, C. 2016, \apj, 824, 41, \dodoi{10.3847/0004-637X/824/1/41}

\bibitem[{{Indriolo} \& {McCall}(2012)}]{2012ApJ...745...91I}
{Indriolo}, N., \& {McCall}, B.~J. 2012, \apj, 745, 91,
  \dodoi{10.1088/0004-637X/745/1/91}

\bibitem[{{Keller} \& {Kruijssen}(2022)}]{2022MNRAS.512..199K}
{Keller}, B.~W., \& {Kruijssen}, J.~M.~D. 2022, \mnras, 512, 199,
  \dodoi{10.1093/mnras/stac511}

\bibitem[{{Kim} \& {Ostriker}(2015)}]{2015ApJ...802...99K}
{Kim}, C.-G., \& {Ostriker}, E.~C. 2015, \apj, 802, 99,
  \dodoi{10.1088/0004-637X/802/2/99}

\bibitem[{{Kim} \& {Ostriker}(2017)}]{2017ApJ...846..133K}
---. 2017, \apj, 846, 133, \dodoi{10.3847/1538-4357/aa8599}

\bibitem[{{Kim} \& {Ostriker}(2018)}]{2018ApJ...853..173K}
---. 2018, \apj, 853, 173, \dodoi{10.3847/1538-4357/aaa5ff}

\bibitem[{{Kim} {et~al.}(2017){Kim}, {Ostriker}, \&
  {Raileanu}}]{2017ApJ...834...25K}
{Kim}, C.-G., {Ostriker}, E.~C., \& {Raileanu}, R. 2017, \apj, 834, 25,
  \dodoi{10.3847/1538-4357/834/1/25}

\bibitem[{{Kim} {et~al.}(2020){Kim}, {Ostriker}, {Somerville}, {Bryan},
  {Fielding}, {Forbes}, {Hayward}, {Hernquist}, \&
  {Pandya}}]{2020ApJ...900...61K}
{Kim}, C.-G., {Ostriker}, E.~C., {Somerville}, R.~S., {et~al.} 2020, \apj, 900,
  61, \dodoi{10.3847/1538-4357/aba962}

\bibitem[{{Kim} {et~al.}(2016){Kim}, {Agertz}, {Teyssier}, {Butler},
  {Ceverino}, {Choi}, {Feldmann}, {Keller}, {Lupi}, {Quinn}, {Revaz},
  {Wallace}, {Gnedin}, {Leitner}, {Shen}, {Smith}, {Thompson}, {Turk}, {Abel},
  {Arraki}, {Benincasa}, {Chakrabarti}, {DeGraf}, {Dekel}, {Goldbaum},
  {Hopkins}, {Hummels}, {Klypin}, {Li}, {Madau}, {Mandelker}, {Mayer},
  {Nagamine}, {Nickerson}, {O'Shea}, {Primack}, {Roca-F{\`a}brega}, {Semenov},
  {Shimizu}, {Simpson}, {Todoroki}, {Wadsley}, {Wise}, \& {AGORA
  Collaboration}}]{2016ApJ...833..202K}
{Kim}, J.-h., {Agertz}, O., {Teyssier}, R., {et~al.} 2016, \apj, 833, 202,
  \dodoi{10.3847/1538-4357/833/2/202}

\bibitem[{{Kimm} {et~al.}(2015){Kimm}, {Cen}, {Devriendt}, {Dubois}, \&
  {Slyz}}]{2015MNRAS.451.2900K}
{Kimm}, T., {Cen}, R., {Devriendt}, J., {Dubois}, Y., \& {Slyz}, A. 2015,
  \mnras, 451, 2900, \dodoi{10.1093/mnras/stv1211}

\bibitem[{{Kretschmer} \& {Teyssier}(2020)}]{2020MNRAS.492.1385K}
{Kretschmer}, M., \& {Teyssier}, R. 2020, \mnras, 492, 1385,
  \dodoi{10.1093/mnras/stz3495}

\bibitem[{{Krumholz} \& {McKee}(2005)}]{2005ApJ...630..250K}
{Krumholz}, M.~R., \& {McKee}, C.~F. 2005, \apj, 630, 250,
  \dodoi{10.1086/431734}

\bibitem[{{Lah{\'e}n} {et~al.}(2019){Lah{\'e}n}, {Naab}, {Johansson},
  {Elmegreen}, {Hu}, \& {Walch}}]{2019ApJ...879L..18L}
{Lah{\'e}n}, N., {Naab}, T., {Johansson}, P.~H., {et~al.} 2019, \apjl, 879,
  L18, \dodoi{10.3847/2041-8213/ab2a13}

\bibitem[{{Lah{\'e}n} {et~al.}(2020){Lah{\'e}n}, {Naab}, {Johansson},
  {Elmegreen}, {Hu}, {Walch}, {Steinwandel}, \& {Moster}}]{2020ApJ...891....2L}
---. 2020, \apj, 891, 2, \dodoi{10.3847/1538-4357/ab7190}

\bibitem[{{Leaman} {et~al.}(2012){Leaman}, {Venn}, {Brooks}, {Battaglia},
  {Cole}, {Ibata}, {Irwin}, {McConnachie}, {Mendel}, \&
  {Tolstoy}}]{2012ApJ...750...33L}
{Leaman}, R., {Venn}, K.~A., {Brooks}, A.~M., {et~al.} 2012, \apj, 750, 33,
  \dodoi{10.1088/0004-637X/750/1/33}

\bibitem[{{Li} {et~al.}(2017){Li}, {Bryan}, \&
  {Ostriker}}]{2017ApJ...841..101L}
{Li}, M., {Bryan}, G.~L., \& {Ostriker}, J.~P. 2017, \apj, 841, 101,
  \dodoi{10.3847/1538-4357/aa7263}

\bibitem[{{Lucy}(1977)}]{1977AJ.....82.1013L}
{Lucy}, L.~B. 1977, \aj, 82, 1013, \dodoi{10.1086/112164}

\bibitem[{{Marinacci} {et~al.}(2019){Marinacci}, {Sales}, {Vogelsberger},
  {Torrey}, \& {Springel}}]{2019MNRAS.489.4233M}
{Marinacci}, F., {Sales}, L.~V., {Vogelsberger}, M., {Torrey}, P., \&
  {Springel}, V. 2019, \mnras, 489, 4233, \dodoi{10.1093/mnras/stz2391}

\bibitem[{{Martizzi} {et~al.}(2016){Martizzi}, {Fielding},
  {Faucher-Gigu{\`e}re}, \& {Quataert}}]{2016MNRAS.459.2311M}
{Martizzi}, D., {Fielding}, D., {Faucher-Gigu{\`e}re}, C.-A., \& {Quataert}, E.
  2016, \mnras, 459, 2311, \dodoi{10.1093/mnras/stw745}

\bibitem[{{McQuinn} {et~al.}(2019){McQuinn}, {van Zee}, \&
  {Skillman}}]{2019ApJ...886...74M}
{McQuinn}, K. B.~W., {van Zee}, L., \& {Skillman}, E.~D. 2019, \apj, 886, 74,
  \dodoi{10.3847/1538-4357/ab4c37}

\bibitem[{{Mondal} {et~al.}(2018){Mondal}, {Subramaniam}, \&
  {George}}]{2018AJ....156..109M}
{Mondal}, C., {Subramaniam}, A., \& {George}, K. 2018, \aj, 156, 109,
  \dodoi{10.3847/1538-3881/aad4f6}

\bibitem[{{Naab} \& {Ostriker}(2017)}]{2017ARA&A..55...59N}
{Naab}, T., \& {Ostriker}, J.~P. 2017, \araa, 55, 59,
  \dodoi{10.1146/annurev-astro-081913-040019}

\bibitem[{{Navarro} {et~al.}(1997){Navarro}, {Frenk}, \&
  {White}}]{1997ApJ...490..493N}
{Navarro}, J.~F., {Frenk}, C.~S., \& {White}, S.~D.~M. 1997, \apj, 490, 493,
  \dodoi{10.1086/304888}

\bibitem[{{Orr} {et~al.}(2022){Orr}, {Fielding}, {Hayward}, \&
  {Burkhart}}]{2022ApJ...932...88O}
{Orr}, M.~E., {Fielding}, D.~B., {Hayward}, C.~C., \& {Burkhart}, B. 2022,
  \apj, 932, 88, \dodoi{10.3847/1538-4357/ac6c26}

\bibitem[{{Padoan} \& {Nordlund}(2011)}]{2011ApJ...730...40P}
{Padoan}, P., \& {Nordlund}, {\r{A}}. 2011, \apj, 730, 40,
  \dodoi{10.1088/0004-637X/730/1/40}

\bibitem[{{Pakmor} {et~al.}(2016){Pakmor}, {Springel}, {Bauer}, {Mocz},
  {Munoz}, {Ohlmann}, {Schaal}, \& {Zhu}}]{2016MNRAS.455.1134P}
{Pakmor}, R., {Springel}, V., {Bauer}, A., {et~al.} 2016, \mnras, 455, 1134,
  \dodoi{10.1093/mnras/stv2380}

\bibitem[{{Price}(2008)}]{2008JCoPh.22710040P}
{Price}, D.~J. 2008, Journal of Computational Physics, 227, 10040,
  \dodoi{10.1016/j.jcp.2008.08.011}

\bibitem[{{Rantala} {et~al.}(2021){Rantala}, {Naab}, \&
  {Springel}}]{2021MNRAS.502.5546R}
{Rantala}, A., {Naab}, T., \& {Springel}, V. 2021, \mnras, 502, 5546,
  \dodoi{10.1093/mnras/stab057}

\bibitem[{{Rathjen} {et~al.}(2021){Rathjen}, {Naab}, {Girichidis}, {Walch},
  {W{\"u}nsch}, {Dinnbier}, {Seifried}, {Klessen}, \&
  {Glover}}]{2021MNRAS.504.1039R}
{Rathjen}, T.-E., {Naab}, T., {Girichidis}, P., {et~al.} 2021, \mnras, 504,
  1039, \dodoi{10.1093/mnras/stab900}

\bibitem[{{R{\'e}my-Ruyer} {et~al.}(2014){R{\'e}my-Ruyer}, {Madden},
  {Galliano}, {Galametz}, {Takeuchi}, {Asano}, {Zhukovska}, {Lebouteiller},
  {Cormier}, {Jones}, {Bocchio}, {Baes}, {Bendo}, {Boquien}, {Boselli},
  {DeLooze}, {Doublier-Pritchard}, {Hughes}, {Karczewski}, \&
  {Spinoglio}}]{2014A&A...563A..31R}
{R{\'e}my-Ruyer}, A., {Madden}, S.~C., {Galliano}, F., {et~al.} 2014, \aap,
  563, A31, \dodoi{10.1051/0004-6361/201322803}

\bibitem[{{Rubio} {et~al.}(2015){Rubio}, {Elmegreen}, {Hunter}, {Brinks},
  {Cort{\'e}s}, \& {Cigan}}]{2015Natur.525..218R}
{Rubio}, M., {Elmegreen}, B.~G., {Hunter}, D.~A., {et~al.} 2015, \nat, 525,
  218, \dodoi{10.1038/nature14901}

\bibitem[{{Scannapieco} {et~al.}(2012){Scannapieco}, {Wadepuhl}, {Parry},
  {Navarro}, {Jenkins}, {Springel}, {Teyssier}, {Carlson}, {Couchman}, {Crain},
  {Dalla Vecchia}, {Frenk}, {Kobayashi}, {Monaco}, {Murante}, {Okamoto},
  {Quinn}, {Schaye}, {Stinson}, {Theuns}, {Wadsley}, {White}, \&
  {Woods}}]{2012MNRAS.423.1726S}
{Scannapieco}, C., {Wadepuhl}, M., {Parry}, O.~H., {et~al.} 2012, \mnras, 423,
  1726, \dodoi{10.1111/j.1365-2966.2012.20993.x}

\bibitem[{{Semenov} {et~al.}(2016){Semenov}, {Kravtsov}, \&
  {Gnedin}}]{2016ApJ...826..200S}
{Semenov}, V.~A., {Kravtsov}, A.~V., \& {Gnedin}, N.~Y. 2016, \apj, 826, 200,
  \dodoi{10.3847/0004-637X/826/2/200}

\bibitem[{{Semenov} {et~al.}(2017){Semenov}, {Kravtsov}, \&
  {Gnedin}}]{2017ApJ...845..133S}
---. 2017, \apj, 845, 133, \dodoi{10.3847/1538-4357/aa8096}

\bibitem[{{Simpson} {et~al.}(2015){Simpson}, {Bryan}, {Hummels}, \&
  {Ostriker}}]{2015ApJ...809...69S}
{Simpson}, C.~M., {Bryan}, G.~L., {Hummels}, C., \& {Ostriker}, J.~P. 2015,
  \apj, 809, 69, \dodoi{10.1088/0004-637X/809/1/69}

\bibitem[{{Smith} {et~al.}(2017){Smith}, {Bryan}, {Glover}, {Goldbaum}, {Turk},
  {Regan}, {Wise}, {Schive}, {Abel}, {Emerick}, {O'Shea}, {Anninos}, {Hummels},
  \& {Khochfar}}]{2017MNRAS.466.2217S}
{Smith}, B.~D., {Bryan}, G.~L., {Glover}, S.~C.~O., {et~al.} 2017, \mnras, 466,
  2217, \dodoi{10.1093/mnras/stw3291}

\bibitem[{{Smith}(2021)}]{2021MNRAS.502.5417S}
{Smith}, M.~C. 2021, \mnras, 502, 5417, \dodoi{10.1093/mnras/stab291}

\bibitem[{{Smith} {et~al.}(2021){Smith}, {Bryan}, {Somerville}, {Hu},
  {Teyssier}, {Burkhart}, \& {Hernquist}}]{2021MNRAS.506.3882S}
{Smith}, M.~C., {Bryan}, G.~L., {Somerville}, R.~S., {et~al.} 2021, \mnras,
  506, 3882, \dodoi{10.1093/mnras/stab1896}

\bibitem[{{Smith} {et~al.}(2018){Smith}, {Sijacki}, \&
  {Shen}}]{2018MNRAS.478..302S}
{Smith}, M.~C., {Sijacki}, D., \& {Shen}, S. 2018, \mnras, 478, 302,
  \dodoi{10.1093/mnras/sty994}

\bibitem[{{Smith} {et~al.}(2020){Smith}, {Tre{\ss}}, {Sormani}, {Glover},
  {Klessen}, {Clark}, {Izquierdo}, {Duarte-Cabral}, \&
  {Zucker}}]{2020MNRAS.492.1594S}
{Smith}, R.~J., {Tre{\ss}}, R.~G., {Sormani}, M.~C., {et~al.} 2020, \mnras,
  492, 1594, \dodoi{10.1093/mnras/stz3328}

\bibitem[{{Somerville} \& {Dav{\'e}}(2015)}]{2015ARA&A..53...51S}
{Somerville}, R.~S., \& {Dav{\'e}}, R. 2015, \araa, 53, 51,
  \dodoi{10.1146/annurev-astro-082812-140951}

\bibitem[{{Springel}(2005)}]{2005MNRAS.364.1105S}
{Springel}, V. 2005, \mnras, 364, 1105,
  \dodoi{10.1111/j.1365-2966.2005.09655.x}

\bibitem[{{Springel}(2010)}]{2010MNRAS.401..791S}
---. 2010, \mnras, 401, 791, \dodoi{10.1111/j.1365-2966.2009.15715.x}

\bibitem[{{Springel} {et~al.}(2005){Springel}, {Di Matteo}, \&
  {Hernquist}}]{2005MNRAS.361..776S}
{Springel}, V., {Di Matteo}, T., \& {Hernquist}, L. 2005, \mnras, 361, 776,
  \dodoi{10.1111/j.1365-2966.2005.09238.x}

\bibitem[{{Steinwandel} {et~al.}(2022){Steinwandel}, {Bryan}, {Somerville},
  {Hayward}, \& {Burkhart}}]{2022arXiv220509774S}
{Steinwandel}, U.~P., {Bryan}, G.~L., {Somerville}, R.~S., {Hayward}, C.~C., \&
  {Burkhart}, B. 2022, arXiv e-prints, arXiv:2205.09774.
\newblock \doarXiv{2205.09774}

\bibitem[{{Steinwandel} {et~al.}(2020){Steinwandel}, {Moster}, {Naab}, {Hu}, \&
  {Walch}}]{2020MNRAS.495.1035S}
{Steinwandel}, U.~P., {Moster}, B.~P., {Naab}, T., {Hu}, C.-Y., \& {Walch}, S.
  2020, \mnras, 495, 1035, \dodoi{10.1093/mnras/staa821}

\bibitem[{{Teyssier}(2002)}]{2002A&A...385..337T}
{Teyssier}, R. 2002, \aap, 385, 337, \dodoi{10.1051/0004-6361:20011817}

\bibitem[{{Truelove} {et~al.}(1997){Truelove}, {Klein}, {McKee}, {Holliman},
  {Howell}, \& {Greenough}}]{1997ApJ...489L.179T}
{Truelove}, J.~K., {Klein}, R.~I., {McKee}, C.~F., {et~al.} 1997, \apjl, 489,
  L179, \dodoi{10.1086/310975}

\bibitem[{{Walch} {et~al.}(2015){Walch}, {Girichidis}, {Naab}, {Gatto},
  {Glover}, {W{\"u}nsch}, {Klessen}, {Clark}, {Peters}, {Derigs}, \&
  {Baczynski}}]{2015MNRAS.454..238W}
{Walch}, S., {Girichidis}, P., {Naab}, T., {et~al.} 2015, \mnras, 454, 238,
  \dodoi{10.1093/mnras/stv1975}

\bibitem[{{Wall} {et~al.}(2020){Wall}, {Mac Low}, {McMillan}, {Klessen},
  {Portegies Zwart}, \& {Pellegrino}}]{2020ApJ...904..192W}
{Wall}, J.~E., {Mac Low}, M.-M., {McMillan}, S. L.~W., {et~al.} 2020, \apj,
  904, 192, \dodoi{10.3847/1538-4357/abc011}

\bibitem[{{Weinberger} {et~al.}(2020){Weinberger}, {Springel}, \&
  {Pakmor}}]{2020ApJS..248...32W}
{Weinberger}, R., {Springel}, V., \& {Pakmor}, R. 2020, \apjs, 248, 32,
  \dodoi{10.3847/1538-4365/ab908c}

\bibitem[{{Wheeler} {et~al.}(2019){Wheeler}, {Hopkins}, {Pace},
  {Garrison-Kimmel}, {Boylan-Kolchin}, {Wetzel}, {Bullock}, {Kere{\v{s}}},
  {Faucher-Gigu{\`e}re}, \& {Quataert}}]{2019MNRAS.490.4447W}
{Wheeler}, C., {Hopkins}, P.~F., {Pace}, A.~B., {et~al.} 2019, \mnras, 490,
  4447, \dodoi{10.1093/mnras/stz2887}

\bibitem[{{Wolfire} {et~al.}(2003){Wolfire}, {McKee}, {Hollenbach}, \&
  {Tielens}}]{2003ApJ...587..278W}
{Wolfire}, M.~G., {McKee}, C.~F., {Hollenbach}, D., \& {Tielens}, A.~G.~G.~M.
  2003, \apj, 587, 278, \dodoi{10.1086/368016}

\end{thebibliography}
\bibliographystyle{aasjournal}

\appendix
%\section{Pressure floor}
%\begin{equation}
%u_{\rm fl} = \frac{L_{J,\rm th}^2 G \rho_{\rm gas}}{\gamma (\gamma-1  )\pi}
%\end{equation}
%It helps for Gizmo and Gadget.
%Arepo still blows out.

\section{Photoelectric heating efficiency}\label{app:PEeff}

\begin{figure*}
	\centering
	\includegraphics[trim=2.5cm 0cm 4cm 0cm,clip, width=0.99\linewidth]{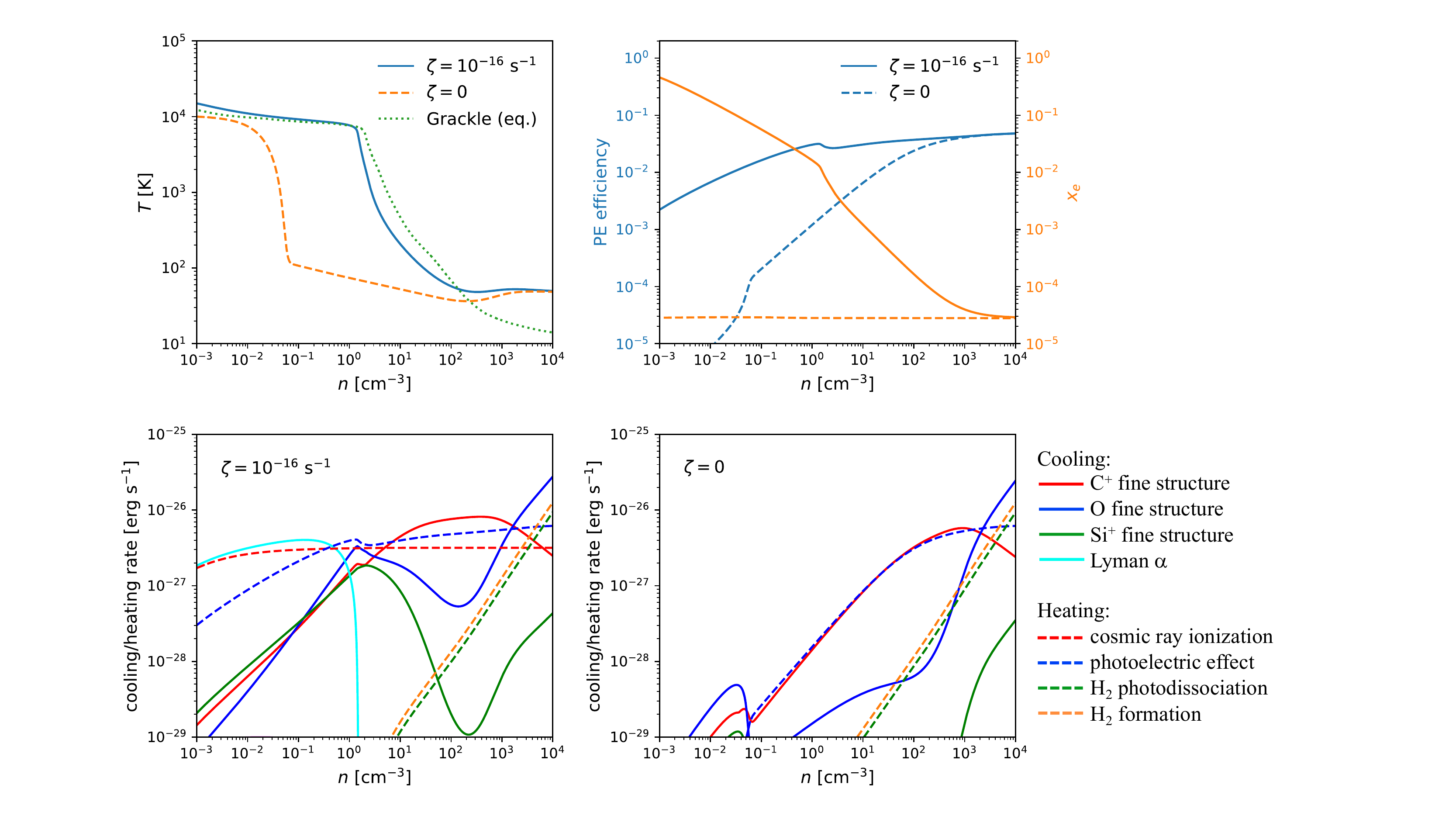}
	\caption{
		\textit{Upper left}: 
		equilibrium temperature as a function of density for the {\sc Grackle} equilibrium module (dotted green),
		{\sc SGchem} (see text) with cosmic rate ionization rate $\zeta = 10^{-16}~{\rm s^{-1}}$ (solid blue),
		and {\sc SGchem} with $\zeta = 0$ (dashed orange).
		\textit{Upper right}:
		photoelectric efficiency ($\epsilon_{\rm PE}$, blue) and electron abundance ($x_{\rm e}$, orange) as a function of density 
		for {\sc SGchem} with $\zeta = 10^{-16}~{\rm s^{-1}}$ (solid) and $\zeta = 0$ (dashed).
		\textit{Lower left}:
		individual cooling and heating processes as a function of density in {\sc SGchem} with $\zeta = 10^{-16}~{\rm s^{-1}}$.
		\textit{Lower right}:
		same as lower left, but with $\zeta = 0$.
		Adopting a constant $\epsilon_{\rm PE}$ is a fair approximation
		of the more realistic situation where both photoelectric effect and cosmic ray ionization are present.		
	}
	\label{fig:cooling_gr_vs_SG}
\end{figure*}

In this section,
we investigate if our adopted constant photoelectric efficiency $\epsilon_{\rm PE} = 0.05$ 
is a good approximation (see Eq.~\ref{eq:PEheat}).
In the upper left panel of Fig.~\ref{fig:cooling_gr_vs_SG},
we show the gas temperature at thermal equilibrium as a function of density 
by running our adopted {\sc Grackle} module for 1 Gyr at each density bin (green dotted line).
The initial temperature is set at $10^4$~K and the metallicity is 0.1 $Z_\odot$.
In comparison,
we also show the equilibrium temperature as a function of density
obtained by
running the non-equilibrium chemistry code developed by Simon Glover \citep{2007ApJS..169..239G,2012MNRAS.421..116G} 
(hereafter {\sc SGchem}) for 1 Gyr at each density bin:
the solid blue line is with a cosmic ray (CR) ionization rate $\zeta = 10^{-16}~{\rm s^{-1}}$ 
motivated by H$^+_3$ observations in the Milky Way \citep{2012ApJ...745...91I}
while the dashed orange line is with $\zeta = 0$.

The photoelectric efficiency adopted in {\sc SGchem} is
\begin{equation}\label{eq:psi}
\epsilon_{\rm PE} = \frac{0.049}{1 + (0.004\psi^{0.73})} 
+ \frac{0.037 (T/10000)^{0.7}}{1 + 2\times 10^{-4} \psi}
\end{equation}
where 
%$\psi = G_{\rm eff} T^{0.5}/n_{\rm e}$
$\psi = G_0 T^{0.5}/n_{\rm e}$
and
$n_{\rm e}$ is the electron number density,
following \citet{1994ApJ...427..822B,2003ApJ...587..278W,2004ApJ...612..921B}.
Here we have assumed that dust shielding is negligible which is a fair approximation at low metallicity.
%$G_{\rm eff} = G_0 {\rm exp}(-1.33\times 10^{-21} Z^\prime_{\rm d} N_{\rm H,tot})$ is the attenuated radiation field strength in units of the Habing field \citep{1968BAN....19..421H}.

We find excellent agreement between {\sc Grackle} and {\sc SGchem} with $\zeta = 10^{-16}~{\rm s^{-1}}$
up to $n \lesssim 10^2~{\rm cm^{-3}}$,
which is reassuring.
%where the main discrepancy at $n > 10^2~{\rm cm^{-3}}$ is due to chemistry heating.
In contrast,
{\sc SGchem} with $\zeta = 0$ leads to a significantly lower equilibrium temperature for a broad range of densities.
This is caused by a severely underestimated electron abundance $x_{\rm e} \equiv n_{\rm e} / n$ when CR ionization is switched off, as shown in the upper right panel of Fig.~\ref{fig:cooling_gr_vs_SG} (shown in orange),
which in turn strongly underestimates $\epsilon_{\rm PE}$ (shown in blue).

The individual cooling and heating processes in {\sc SGchem} are shown 
in the lower left (for $\zeta = 10^{-16}~{\rm s^{-1}}$) and lower right (for $\zeta = 0$) panels.
For $\zeta = 10^{-16}~{\rm s^{-1}}$,
heating is dominated by photoelectric effect at $n > 0.3~{\rm cm^{-3}}$
and by CR ionization at $n < 0.3~{\rm cm^{-3}}$.
The total heating rate is almost constant with $n$,
which is balanced by Lyman alpha cooling at low $n$ and by fine structure line cooling at high $n$.
At $n > 10^3~{\rm cm^{-3}}$,
heating from H$_2$ photodissociation and H$_2$ formation become non-negligible,
which might explain the discrepancy in the equilibrium temperature in this regime.
In contrast,
for $\zeta = 0$,
not only does CR ionization heating vanish by construction,
but the photoelectric heating is also significantly suppressed caused by the artificially low electron abundance.
As a result,
the equilibrium temperature becomes unreasonably low ($T < 10^2$~K) even for typical densities in the diffuse ISM.
This has been seen in
\citet{2017MNRAS.471.2151H}
where the authors adopted {\sc SGchem} but did not include CR ionization.
Interestingly,
when applied to the hydrodynamical simulations of \citet{2017MNRAS.471.2151H},
such low temperatures are never reached
as SN feedback (and the turbulence it drives) 
provides extra heating and ionization,
which has a similar effect as CR ionization.
Therefore,
the resulting phase diagram is quite similar to what we found in Fig.~\ref{fig:pdcodecompare}.
That said, 
having a more realistic thermal balance in a static medium is still desirable.
We recommend future simulations that use {\sc SGchem} 
(or other similar chemistry codes) 
where
$\epsilon_{\rm PE}$ is calculated via Eq.~\ref{eq:psi}
to always include CR ionization.

To conclude,
adopting a constant $\epsilon_{\rm PE}$ is a good approximation
of the more realistic situation where both photoelectric effect and CR ionization are present,
even though we do not explicitly include the latter.
In reality,
CR ionization either dominates heating at low densities
or provides the crucial free electrons that facilitates photoelectric heating at high densities.

\section{Constant star formation timescale}\label{app:tSFR2Gyr}
\begin{figure*}
	\centering
	\includegraphics[trim=1cm 0cm 0.5cm 0cm,clip,width=0.99\linewidth]{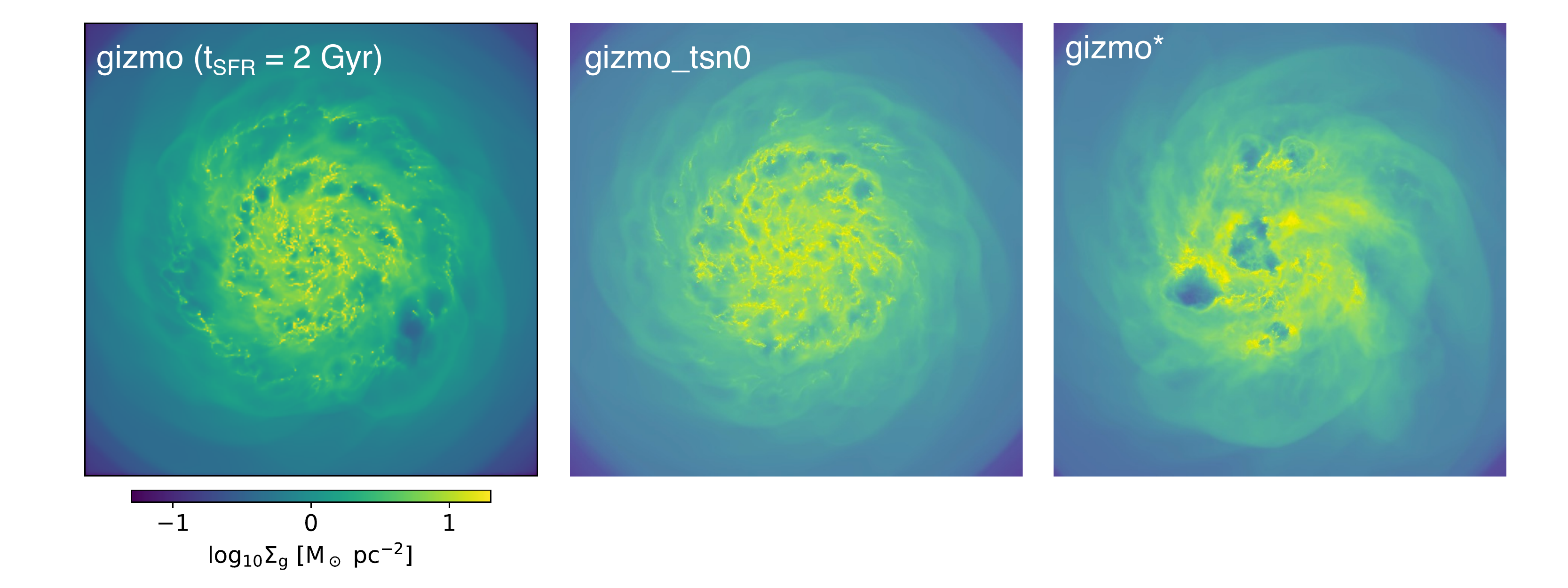}
	\caption{
		{Gas surface density maps of different {\sc Gizmo} models at $t = 500$~Myr.	The constant depletion time model (left) is qualitatively more similar to the \textit{gizmo\_tsn0} model with a smooth gas morphology, as opposed to the \textit{gizmo*} model which shows large SN bubbles.}}
	\label{fig:maps_tSGR2Gyr}
\end{figure*}

\begin{figure*}
	\centering
	\includegraphics[trim=1cm 1cm 0cm 1cm,clip,width=0.99\linewidth]{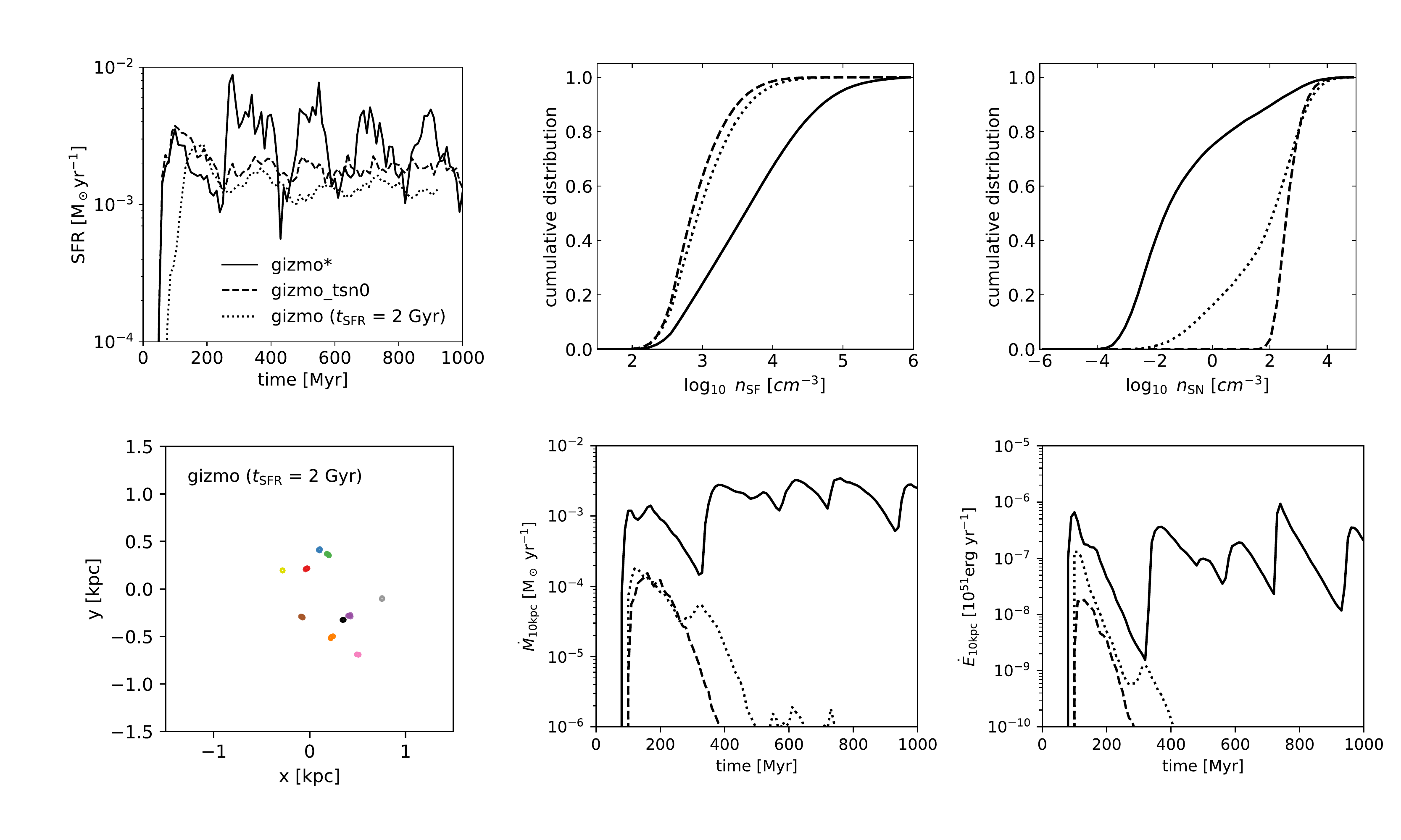}
	\caption{
		{Comparison among \textit{gizmo*} (solid), \textit{gizmo\_tsn0} (dashed), and the constant depletion time model (dotted).
		\textit{Upper left}: SFR as a function of time.
		\textit{Upper middle}: cumulative distribution of $n_{\rm SF}$.
		\textit{Upper right}: cumulative distribution of $n_{\rm SN}$.
		\textit{Lower left}: top ten SN clusters for the constant depletion model only (face-on view).
		\textit{Lower middle}: mass outflow rate as a function of time.
		\textit{Lower right}: energy outflow rate as a function of time. 
		The constant depletion time model behaves more similar to the \textit{gizmo\_tsn0} model -- 
		both have non-bursty star formation and vanishing outflows as a result of low SN clustering.
		}
	}
	\label{fig:tSFR2Gyr_SFR}
\end{figure*}

{In this section, we present a numerical experiment suggested by the anonymous referee that will further strengthen our conclusion on the importance of SN clustering.	
As we discussed in Section~\ref{sec:SNcluster},
the strong SN clustering in Lagrangian codes is due to the locally enhanced SFR from the super-linear density dependence
$\dot{\rho}_{\rm *} = \epsilon_{\rm SF} \rho_{\rm gas} / t_{\rm ff} \propto  \rho_{\rm gas}^{1.5}$.
Here we conduct a simulation using {\sc Gizmo}
with $t_{\rm SN} = 10$~Myr but with the local SFR $\dot{\rho}_{\rm *} = \rho_{\rm gas} / t_{\rm SFR}$
where $t_{\rm SFR} = 2$~Gyr is the local depletion time which is chosen to be a constant.
In other words,
the local SFR has a linear (rather than super-linear) dependence on gas density.
As shown in Fig.~\ref{fig:maps_tSGR2Gyr}, 
this constant depletion time model shows a smooth gas morphology 
similar to the \textit{gizmo\_tsn0} model but with more pronounced dense gas clumps.
This is in contrast to the \textit{gizmo*} model with large SN bubbles.
In addition,
as shown in Fig.~\ref{fig:tSFR2Gyr_SFR}, 
this model shows a smooth, non-bursty SFR as a function of time
and cumulative distributions of $n_{\rm SF}$ and $n_{\rm SN}$ similar to those in the \textit{gizmo\_tsn0} model.
This is because although gas can collapse to densities much higher than the star formation threshold density,
the local SFR is not enhanced leading to a large amount of dense clumps.
These clumps,
instead of quickly convert into stars as in the \textit{gizmo*} model, 
simply wonder around, forming stars at a much lower rate, 
until they get dispersed by SN feedback after 10 Myr.
This leads to low SN clustering,
with the maximum $N_{\rm cl} = 5$ in the entire simulation,
which in turn leads to vanishing outflows.

These results support our argument that the super-linear dependence of the local SFR on density
$\dot{\rho}_{\rm *} \propto \rho_{\rm gas}^{1.5}$ leads to strong SN clustering in Lagrangian codes.
Adopting a constant local depletion time can make Lagrangian codes behave more like the fiducial {\sc Ramses} model.
However, just like our $t_{\rm SN} = 0$ models,
this is merely a numerical experiment as it implies that 
the star formation efficiency decreases as density increases,
which is not physically justified.
}

\end{document}